\documentclass{ws-mpla}
\usepackage{graphicx}
\usepackage[super,compress]{cite}
\usepackage[breaklinks]{hyperref}  
\hypersetup{colorlinks,urlcolor=black,citecolor=black,linkcolor=black,filecolor=black}
\usepackage{breakurl}
\usepackage{color}
\usepackage{amsmath}
\usepackage{slashed}

\def\POWHEGBOX{{\tt POWHEG BOX}}
\def\POWHEG{{\tt POWHEG}}
\def\SHERPA{{\tt SHERPA}}

\def\GoSam{{\tt GoSam}}
\def\OpenLoops{{\tt OpenLoops}}
\def\MCNLO{{\tt MC@NLO}}

\def\MadaMCNLO{{\tt MadGraph5\_aMC@NLO}}
\def\MadGraph{{\tt MadGraph}}
\def\PYTHIA{{\tt PYTHIA}}
\def\HERWIG{{\tt HERWIG}}
\def\MCFM{{\tt MCFM}}
\begin{document}

\thispagestyle{plain}

\def\bib{B\kern-.05em{I}\kern-.025em{B}\kern-.08em}
\def\btex{B\kern-.05em{I}\kern-.025em{B}\kern-.08em\TeX}

\title{{\btex}ing in MPLA {\bib}Style}

\markboth{Febres Cordero, and Reina}
{Vector-boson production in association with $b$ jets at Hadron Colliders}

\title{
\vspace{-1.0cm}
\hbox{\rm\small\textnormal{
FR-PHENO-2015-004}\break}
\hbox{$\null$\break}
\vspace{1.0cm}
ELECTROWEAK GAUGE-BOSON PRODUCTION IN ASSOCIATION WITH B JETS AT HADRON COLLIDERS}

\author{F.~FEBRES CORDERO}
\address{Albert-Ludwigs-Universit\"at Freiburg, Physikalisches Institut \\
D-79104 Freiburg, Germany\\
ffebres@physik.uni-freiburg.de}

\author{L.~REINA}
\address{Department of Physics, Florida State University, \\
Tallahassee, Florida, 32306-4350 USA\\
reina@hep.fsu.edu}

\maketitle

\pub{Received (Day Month Year)}{Revised (Day Month Year)}

\begin{abstract}

  The production of both charged and neutral electroweak gauge bosons
  in association with $b$ jets has attracted a lot of experimental and
  theoretical attention in recent years because of its central role in
  the physics programs of both the Fermilab Tevatron and the CERN
  Large Hadron Collider. The improved level of accuracy achieved both
  in the theoretical predictions and experimental measurements of
  these processes can promote crucial developments in modeling
  $b$-quark jets and $b$-quark parton distribution functions, and can
  provide a more accurate description of some of the most important
  backgrounds to the measurement of Higgs-boson couplings and
  several new physics searches.  In this paper we review the
  status of theoretical predictions for cross sections and kinematic
  distributions of processes in which an electroweak gauge boson is
  produced in association with up to two $b$ jets in hadronic
  collisions, namely $p\bar{p}, pp\rightarrow V+1b$~jet and
  $p\bar{p},pp\rightarrow V+2b$~jets with $V=W^\pm, Z/\gamma^*,
  \gamma$. Available experimental measurements at both the Fermilab
  Tevatron and the CERN Large Hadron Collider are also reviewed and
  their comparison with theoretical predictions is discussed.

\end{abstract}
\ccode{PACS Nos.: 13.85.Ni, 12.38.Bx, 14.65.Fy, 14.80.Bn}

\section{Introduction}
\label{sec:intro}
\noindent

The production of both charged ($W^\pm$) and neutral
($Z/\gamma^*,\gamma$) electroweak gauge bosons in association with $b$
jets (generically denoted as $V+b$-jet production) has attracted a lot
of experimental and theoretical attention in recent years because of
its central role in the physics programs of both the Tevatron and the
Large Hadron Collider (LHC). Many formal developments have improved
the accuracy of theoretical predictions to the level of accuracy
necessary to compare with the results that have meanwhile been
presented by both the Tevatron and the LHC experiments.  In this paper
we present an updated review of this topic and collect the most
important results that can serve as reference for future studies.

From a physics standpoint $W$ and $Z$ production with $b$ jets is
first of all one of the main irreducible backgrounds to the
measurement of a Higgs boson in the $WH$ and $ZH$
associated-production channels, which played a leading role for
Higgs-boson searches at the Tevatron and will be instrumental in
constraining the couplings of the discovered Higgs
boson~\cite{Aad:2012tfa,Chatrchyan:2012ufa} at the LHC. $W+b$~jets is
also and important irreducible background in the measurement of
single-top production, and $Z+b$~jets can mimic the $H\rightarrow
ZZ^*\rightarrow 4l$ ($l$=lepton) final state when associated $B$
mesons decay leptonically.  In general, electroweak vector bosons
(namely $W^\pm$ and $Z$) and $b$ quarks often appear in the decay
chain of heavy massive particles in models beyond the Standard Model
(SM), making of the $V+b$-jet processes important backgrounds for
searches of new physics at the LHC.

At the same time, the production of a SM vector boson ($W^\pm$,
$Z/\gamma^*$, and $\gamma$) with heavy quarks (both $b$ and $c$) is
per se an important test of QCD for two main reasons. The first reason
is that it allows to study the dynamics of $b$ (and $c$) jets and
improve their modeling in many other more challenging SM processes
that represent important backgrounds to new physics searches (consider
for example $t\bar{t}+b$ jets).  The second important reason is that
the production of $W^\pm$, $Z/\gamma^*$, and $\gamma$ with heavy
quarks ($b$ and $c$) offers a unique possibility to measure the
bottom- and charm-quark parton distribution functions (PDF) using only
collider data, as will become clearer (for the $b$ case) from a closer
inspection of the parton-level processes that are at the core of
$V+b$-jet production.

\begin{figure}[ht]
\begin{center}
\includegraphics[scale=0.7]{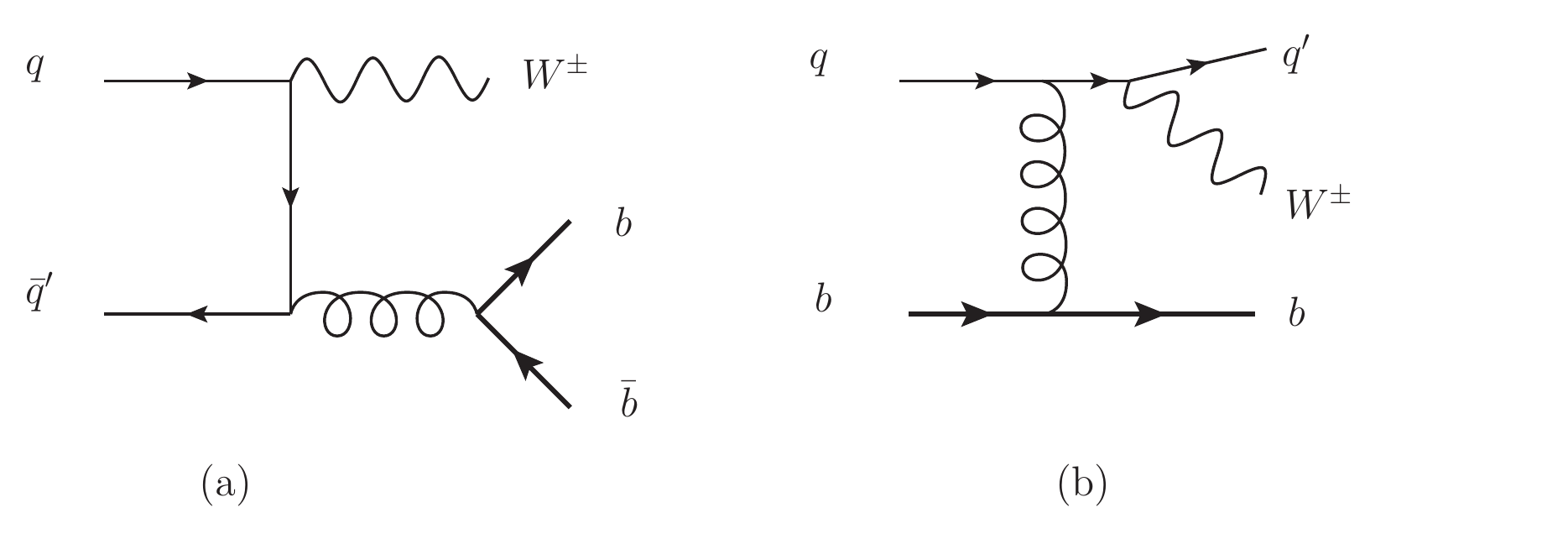}
\caption{Examples of tree-level Feynman diagrams contributing to
  $W^\pm+b$ signatures at hadron colliders, without and with
  contributions from initial-state $b$ quarks.}
\label{fig:diags-Wbb}
\end{center}
\end{figure}

\begin{figure}[ht]
\begin{center}
\includegraphics[scale=0.7]{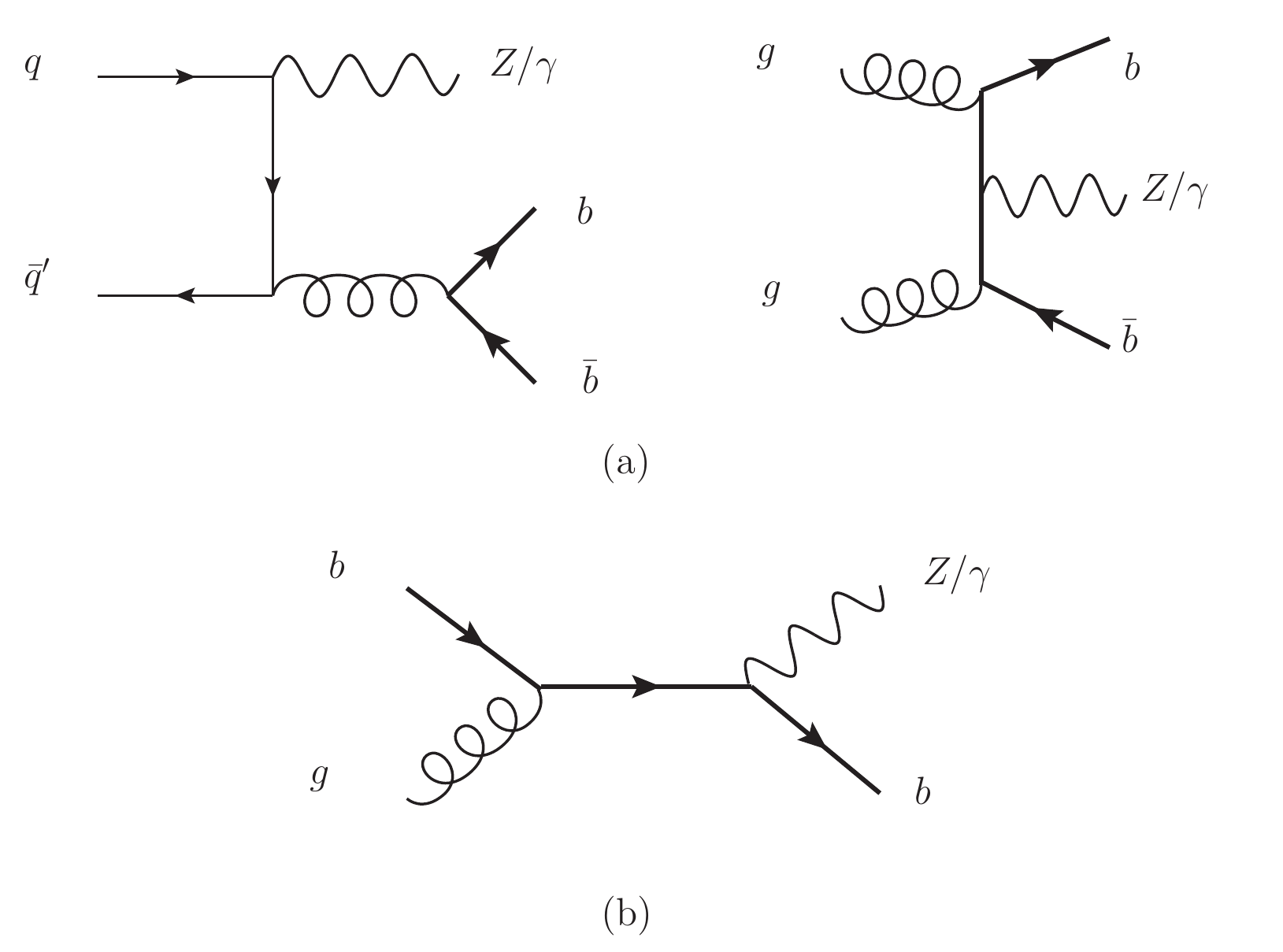}
\caption{Examples of tree-level Feynman diagrams contributing to
  $Z/\gamma\ b$ signatures at hadron colliders, without and with
  contributions from initial-state $b$ quarks.}
\label{fig:diags-Zbb}
\end{center}
\end{figure}

The parton-level structure of $V+b$~jets differs for charged
($V=W^\pm$) and neutral ($V=Z/\gamma^*,\gamma$) gauge bosons.
Furthermore, such structure depends on the number of light flavors
($N_f=4$ or $N_f=5$) that are allowed in the initial-state hadrons
($p$ and $\bar{p}$).  Given the non-negligible mass of the bottom
quark ($m_b\gg\Lambda_{QCD}$), the question of considering a massless
bottom quark with an initial-state parton density in the (anti)proton
is non trivial and is intuitively related to the energy scale of the
process considered. Since typically $V+b$-jet processes involve a
range of energies from few tens to few hundreds of GeV, the answer is
not clear cut and the problem cannot be addressed without considering
the perturbative QCD structure of the corresponding physical
observables (total and differential cross sections), as we will
discuss in more detail in Section~\ref{sec:theory}.  Considering only
the first-order (or tree-level) processes, let us remind the reader
that $W^\pm+2b$~jets can proceed only through
$q\bar{q}^\prime\rightarrow W^\pm b\bar{b}$ (see
Figure~\ref{fig:diags-Wbb}.a), while $W^\pm+1b$~jet can proceed
through the same channel if we assume $N_f=4$ and also through
$bq\rightarrow W^\pm b+q^\prime$ (see Figure~\ref{fig:diags-Wbb}.b) if
we assume $N_f=5$. Similarly, $Z/\gamma+2b$~jets can only proceed (at
lowest order in $\alpha_s$) through $q\bar{q}\rightarrow Z/\gamma
b\bar{b}$ and $gg\rightarrow Z/\gamma b\bar{b}$ (see
Figure~\ref{fig:diags-Zbb}.a), while $Z/\gamma+1b$~jet can proceed
through the same processes if we assume $N_f=4$, or through
$bg\rightarrow bZ/\gamma$ if we assume $N_f=5$ (see
Figure~\ref{fig:diags-Zbb}.b). Notice that in this discussion we keep
only the Cabibbo sector of the flavor-mixing CKM matrix and neglect  
processes initiated by two $b$ (anti)quarks (like
e.g. $b\bar{b}\rightarrow \gamma b\bar b$). These approximations have
effects at the 1\% level in the observables studied. The diagrams in
Figures~\ref{fig:diags-Wbb}.b and \ref{fig:diags-Zbb}.b clearly
illustrate the possibility of measuring the $b$-quark PDF by measuring
$V+b$-jet processes.

On the theoretical side, total and differential hadronic cross
sections including Next-to-Leading Order (NLO) QCD corrections have
been first calculated for $W^\pm b\bar{b}$ and $Zb\bar{b}$ production
in the limit of zero bottom-quark
mass,\cite{Ellis:1998fv,Campbell:2000bg} followed more recently by
fully massive calculations ($m_b\neq 0$) for $W^\pm
b\bar{b}$,\cite{FebresCordero:2006sj,Cordero:2009kv,Badger:2010mg}
$Zb\bar{b}$,\cite{FebresCordero:2008ci,Cordero:2009kv} and $\gamma
b\bar{b}$\cite{Hartanto:2013aha} production. The NLO QCD calculations
of $W^\pm b\bar{b}$ hadronic production (for $m_b\neq 0$), including
spin-correlation effects from $W\rightarrow l\nu_l$
decay,\cite{Badger:2010mg} and of $Z/\gamma^*b\bar{b}$ (with $m_b=0$
and $Z/\gamma^*\rightarrow l^+l^-$) are available in
\MCFM.\cite{mcfm7} These NLO QCD corrections can now also be
calculated with several automated tools like for example
\MadaMCNLO\cite{Alwall:2014hca}, and \SHERPA\cite{Gleisberg:2008ta} in
conjunction with \GoSam\cite{Cullen:2014yla} or
\OpenLoops\cite{Cascioli:2011va}. A study of $W^\pm b\bar{b}$, and
$Z/\gamma^* b\bar{b}$ production at hadron colliders, including NLO
QCD corrections, $m_b$ effects, and spin-correlation effects from $W$
and $Z$ decays, as well as the interface with parton-shower Monte
Carlo generators like
\PYTHIA{}~\cite{Sjostrand:2006za,Sjostrand:2007gs} and
\HERWIG{}~\cite{Marchesini:1991ch,Corcella:2000bw} via the \MCNLO{}
method,\cite{Frixione:2002ik,Frixione:2003ei} has been presented in
Ref.~\refcite{Frederix:2011qg}.  The NLO QCD calculation of
$pp,p\bar{p}\rightarrow W^\pm b\bar{b}$ from
Refs.~\refcite{FebresCordero:2006sj} and \refcite{Cordero:2009kv} has
also been interfaced with Pythia and Herwig using the \POWHEG{}
method,\cite{Nason:2004rx,Frixione:2007vw} and is available in the
\POWHEGBOX{} framework\cite{Oleari:2011ey}. Recently the NLO QCD
hadronic cross section for $W^\pm b\bar{b}+j$ has been implemented in
the \POWHEGBOX{}\cite{Luisoni:2015mpa} (see also the calculation of
the corresponding $O(\alpha_s)$ virtual corrections presented in
Ref.~\refcite{Reina:2011mb}).

Higher-order QCD corrections are also known for the $N_f=5$ processes.
In particular the first complete calculation of $W^\pm$ hadronic
production with at least one $b$ jet ($pp,p\bar{p}\rightarrow W+b+X$),
including NLO QCD corrections and full $b$-mass effects, has been
presented in Ref.~\refcite{Campbell:2008hh}, updated in
Ref.~\refcite{Caola:2011pz}, and is now available through
\MCFM.\cite{mcfm7} The calculation was based on separate contributions
from Refs.~\refcite{Campbell:2006cu}, \refcite{FebresCordero:2006sj},
\refcite{Cordero:2009kv}, and \refcite{Badger:2010mg}.  NLO QCD
corrections to $Z$ hadronic production with one $b$ jet
($pp,p\bar{p}\rightarrow Z+b$) and with at least one $b$ jet
($pp,p\bar{p}\rightarrow Z+b+X$) have been calculated in
Refs.~\refcite{Campbell:2003dd} and \refcite{Campbell:2005zv}
respectively, and are available through \MCFM. Finally, the hadronic
production of a photon and one $b$ jet ($pp,p\bar{p}\rightarrow
\gamma+b$) at NLO QCD has been calculated in
Ref.~\refcite{Stavreva:2009vi} and independently cross-checked in
Ref.~\refcite{Hartanto:2013aha}. Some of these processes are now being
computed using automated NLO QCD parton-shower Monte Carlos like
e.g. \MadaMCNLO,\footnote{ Results from \MadaMCNLO{} have been
  published in ATLAS and CMS papers as we will summarize in
  Section~\ref{sec:results}} although some aspects of the
parton-shower generation in the $N_f=5$ scheme still need to be
studied, as we will discuss in Section~\ref{sec:theory}.

On the experimental side, both the CDF and D0 experiments at the
Tevatron as well as the ATLAS and CMS experiments at the LHC have
provided a variety of results for $V+1b$-jet and $V+2b$-jet
production, both with and without extra light jets. At the Tevatron,
the cross section for $W^\pm+1b$~jet has been measured by both
CDF~\cite{Aaltonen:2009qi} and D0.\cite{D0:2012qt,Abazov:2014fka} CDF
has also measured the cross section for
$Z+1b$~jet,\cite{Aaltonen:2008mt} and D0 the ratio of the cross
sections for $Z+b$~jets and $Z+j$ ($j$=light jets). The production of
\cite{Abazov:2013uza} $\gamma+1b$ jet has further been measured by
CDF~\cite{Aaltonen:2013coa} and D0,\cite{Abazov:2012ea,Abazov:2014hoa}
as well as of $\gamma+2b$ jets by D0.\cite{Abazov:2014hoa} At the LHC,
we now have measurements of $W^\pm+1b$ jet and $W^\pm+2b$ jets from
ATLAS,\cite{Aad:2011kp,Aad:2013vka} and of $W^\pm+2b$~jets from
CMS\cite{Chatrchyan:2013uza}, as well as of $Z+1b$~jet and $Z+2b$~jets
from both ATLAS~\cite{Aad:2011jn,Aad:2014dvb} and
CMS.\cite{Chatrchyan:2012vr,Chatrchyan:2013zja,Chatrchyan:2014dha} We
will summarize the most relevant aspects of these experimental studies
including the exact definition of the signal signatures in
Section~\ref{sec:results}.

The rest of the paper is organized as follows. We briefly review in
Section~\ref{sec:theory} the main structure of the $V+b$-jet cross
sections, including the first order of QCD corrections, for both
$N_f=4$ and $N_f=5$. In Section~\ref{sec:theory} we also discuss the
uncertainties associated with the theoretical predictions for these
processes, and present more specific theoretical issues, like photon
isolation, which is relevant for $\gamma+b$-jet production. A
selection of the most recent experimental results will be presented in
Section~\ref{sec:results} with particular emphasis on the comparison
between theory and experiments as presented in each of the selected
measurements. Section~\ref{sec:conclusions} contains our conclusions
and outlook.

\section{Theoretical Framework}
\label{sec:theory}

In this Section we review the most important aspects of the
calculation of the hadronic cross section for $V+1b$ jet and
$V+2b$ jet production ($V=W^\pm,Z/\gamma^*,\gamma$) at NLO in
QCD. In particular we focus on the meaning of presenting results in
the four- ($N_f=4$, denoted as 4F) and five-flavor ($N_f=5$,
denoted as 5F) schemes, on the matching of the
corresponding parton-level calculations with existing parton-shower
Monte Carlo generators, and on the assessment of the theoretical
uncertainties associated with different stages of the calculation. We
will substantiate our discussion with results taken from various
theoretical studies presented in the literature, and refer to the
corresponding sources for further details.  Several important aspects
of existing theoretical studies are illustrated using the experimental
results collected in Section~\ref{sec:results}.
\subsection{Four- and five-flavor schemes}
\label{subsec:4fns-5fns}

Four- and five-flavor schemes differ by the number of partons that are
allowed to have a non-zero parton distribution function in the
colliding hadrons. Since the bottom quark ($m_b\gg\Lambda_{QCD}$)
lives in the perturbative regime of QCD, the issue of defining a
bottom-quark parton density is non trivial, and is intuitively related
to the energy scale of the process considered. It is important to
understand the relation between the two schemes and emphasize that
they are just different prescriptions to calculate the same physical
processes ($V+b$-jet hadronic production in our case) in perturbative
QCD.  One can indeed tailor the arrangement of the perturbative
expansion of the cross sections to specific energy regimes, where some
classes of corrections may become more important than others. Since 4F
and 5F correspond to different rearrangements of the perturbative
expansion of the cross section for a given process, the two
prescriptions share more terms the higher the perturbative order, and
in some cases significant differences can still appear at NLO in
QCD. A recent systematic study of 4F versus 5F schemes has been
presented in Ref.~\refcite{Maltoni:2012pa}, to which we refer for more
details.

$V+b$-jet hadronic processes offer indeed a broad range of
examples and very effectively lend themselves to illustrate the range
of validity of fixed-order calculations using either the 4F or the 5F
approaches.  Given the multi-scale nature of $V+b$-jet processes, the
accurate theoretical description of these processes requires to
consider the perturbative QCD structure of the corresponding physical
observables (total and differential cross sections) in detail. We will
return to this point with specific examples after having defined more
in detail how 4F and 5F schemes are implemented in the actual
calculation of $V+b$-jet hadronic production.

In the 4F scheme, the initial-state quarks are constrained to be the
four lightest quarks (taken to be massless), there are no bottom
quarks in the initial state, and the final-state bottom quarks are
considered massive. Both $V+2b$-jet and $V+1b$-jet processes receive
contributions at lowest order in QCD from $q\bar{q}$ and $gg$
initiated subprocesses, specifically from $q\bar{q}^\prime\rightarrow
W^\pm b\bar{b}$ in the case of $W^\pm$, and $q\bar{q}\rightarrow
Z/\gamma\, b\bar{b}$ and $gg \rightarrow Z/\gamma\, b\bar{b}$ in the
case of $Z$ and $\gamma$.  Cross sections are computed at fixed order
in QCD (LO, NLO, etc.) and the number of $b$ jets in the final state
is selected applying specific kinematic cuts on the transverse
momentum ($p_T^{b\mathrm{\, jet}}\!>\!p_T^{min}$) and, possibly, the
rapidity/pseudorapidity ($|y^{b\mathrm{\, jet}}|\!<\!y^{max}$ or
$|\eta^{b\mathrm{\, jet}}|\!<\!\eta^{max}$) of the final-state $b$
jets, where it is understood that a given jet algorithm is in place to
select the final-state $b$ jets and light jets.\footnote{We notice
  that in a parton-level calculation, even at NLO in QCD, these jets
  have very limited structure.} At NLO in QCD the cross section for
$Vb\bar{b}$ production receives contributions from both the lowest
order processes listed above and the corresponding $O(\alpha_s)$
virtual and real corrections, more specifically:
\begin{itemlist}
\item one-loop \textit{virtual} corrections to $q\bar{q}^\prime\rightarrow W^\pm
  b\bar{b}$ and $q\bar{q},gg\rightarrow Z/\gamma b\bar{b}$, respectively;
\item one additional parton \textit{real} emission processes,
  i.e. $q\bar{q}^\prime \rightarrow W^\pm b\bar{b}+g$ and
 $qg\rightarrow W^\pm b\bar{b}+q^\prime$ in the case of $W^\pm
 b\bar{b}$ production; as well as
$q\bar{q},gg\rightarrow Z/\gamma b\bar{b}+g$ and
$qg\rightarrow Z/\gamma b\bar{b}+q$ in the case of  $Z/\gamma
b\bar{b}$ production (notice that for convenience we denote by $qg$ also the
  corresponding $\bar{q}g$ channels).
\end{itemlist}

In the 5F scheme, the definition of a bottom-quark PDF arises because
the integration over the phase space of non-identified final-state
bottom quarks (as in the case of $V+1b$-jet processes) gives origin
to potentially large logarithms of the form:
\begin{equation}
\label{eq:lambda_b}
\Lambda_b \equiv \ln\left(\frac{\mu^2}{m_b^2}\right)\,\,\,,
\end{equation}
where $\mu$ is a scale related to the upper bound on the integration
over the transverse momentum, $p_T$, of the final-state bottom
quark,\footnote{The symbol $\mu$ is often used to suggest the
  possibility of setting this upper bound to be the factorization
  scale of the bottom-quark PDF to be defined in the next paragraph.}
while the dependence on $m_b$ comes from the limit of producing a $b$
quark with no transverse momentum.  This happens when a final-state
bottom-quark pair is produced via $g\rightarrow b\bar{b}$ splitting of
an initial-state gluon, and corresponds to the collinear singularity
that would be present in the case of a gluon splitting into massless
quarks ($m_b\rightarrow 0$).  For scales $\mu$ of $\mathcal{O}(M_V)$
or larger these logarithms can be quite large.  Additionally, the same
logarithms appear at every order in the perturbative expansion of the
cross section in the strong coupling, $\alpha_s$, due to recursive
gluon emission from internal bottom-quark lines, as well as virtual
corrections.  These logarithms could hinder the convergence of the
perturbative expansion of the total and differential cross sections.
In the 5F the convergence is improved by introducing a
perturbatively-defined bottom-quark
PDF,\cite{Barnett:1988,Olness:1988,Dicus:1989} defined at lowest order
in $\alpha_s$ as,
\begin{equation}
\label{eq:b_pdf_lo}
\tilde{b}(x,\mu)=\frac{\alpha_s(\mu)}{2\pi}\Lambda_b \int_x^1
\frac{dy}{y} P_{qg}\left(\frac{x}{y}\right)g(y,\mu)\,\,\,,
\end{equation}
\noindent 
where $g(y,\mu)$ is the gluon PDF and $P_{qg}$ is the
Altarelli-Parisi splitting function for $g\rightarrow q\bar{q}$ given
by,
\begin{equation}
\label{eq:ap_qg}
P_{qg} = \frac{1}{2}[z^2 + (1-z)^2]\,\,\,.
\end{equation}
The $\Lambda_b$ logarithms are subsequently resummed through the DGLAP
evolution
equation~\cite{Gribov:1972,Altarelli:1977zs,Dokshitzer:1977}, such
that contributions proportional to $(\alpha_s\Lambda_b)^n$ can be
absorbed, to all orders in $n$, into a leading-logarithm bottom-quark
PDF, while subleading logarithms are recursively resummed when
higher-order corrections are considered in the DGLAP equation.  With
the use of a bottom-quark PDF, the 5F effectively reorders the
perturbative expansion to be one in $\alpha_s$ {\it{and}} $\Lambda_b$,
or $\alpha_s\Lambda_b$. In those kinematic regimes where
$\alpha_s\Lambda_b$ is not a small expansion parameter, the use of a
bottom-quark PDF should improve the stability of the total and
differential cross sections. As observed in
Ref.~\refcite{Maltoni:2012pa}, however, the argument of the
$\Lambda_b$ logarithms, as it arises from the collinear limit of
$g\rightarrow b\bar{b}$ in 4F processes, is often not a constant, but
contains some dependence on the momentum fraction carried by the gluon
($y$ in Eq.~\ref{eq:b_pdf_lo}), which can considerably reduce the
relevance of such logarithmic terms at each perturbative
order. General arguments to estimate their relevance at all
perturbative orders are given for $2\rightarrow 1$ and $2\rightarrow
2$ processes in Ref.~\refcite{Maltoni:2012pa}.

In the 5F, while the production of $V+2b$ jets mainly receives
contributions from light-quark- and gluon-initiated processes
(processes like $b\bar{b}\rightarrow Zb\bar{b}$ are negligible), the
case of $W^\pm +1b$ jet and $Z/\gamma+1b$ jet have rather different
characteristics.  $Z/\gamma+1b$ jet starts at LO with $bg\rightarrow
Z/\gamma \,b$, where, according to our previous discussion, the
initial state $b$-quark density has been introduced to resum the large
logarithms arising in the calculation of $gg\rightarrow Z/\gamma\,
b\bar{b}$, when one of the $b$ quarks in the $g\rightarrow b\bar{b}$
splitting gives origin to a $b$ jet that is not tagged. At lowest
order it corresponds to $\tilde{b}$ in Eq.~(\ref{eq:b_pdf_lo}). At NLO
the following $O(\alpha_s)$ corrections are then to be added:
\begin{itemlist}
\item  one-loop \textit{virtual} corrections to 
$bg\rightarrow Z/\gamma\, b$ (notice that for convenience when we
write $b(q)$ in the initial state we mean generically 
processes initiated by both $b(q)$ and ${\bar b}({\bar q})$);
\item one additional parton \textit{real} emission processes:
$bg\rightarrow Z/\gamma\, b+g$, 
$bq \rightarrow Z/\gamma\, b+q$, $q\bar{q}\rightarrow Z/\gamma
  b\bar{b}$, and $gg\rightarrow Z/\gamma
  b\bar{b}$.
\end{itemlist}
On the other hand, in $W^\pm +1b$-jet production the first
$b$-initiated subprocess starts as a $2\rightarrow 3$ process
($bq\rightarrow W^\pm bq^\prime$) because the first gluon-initiated
subprocess is higher order ($qg\rightarrow W^\pm b\bar{b}+q^\prime$).
Therefore, there are two competing LO subprocess, namely
$q\bar{q}^\prime\rightarrow W^\pm b\bar{b}$ and $bq\rightarrow W^\pm
bq^\prime$ and the NLO QCD cross section includes the following
$O(\alpha_s)$ corrections:
\begin{itemlist}
\item one-loop \textit{virtual} corrections to $q\bar{q}^\prime\rightarrow W^\pm
  b\bar{b}$;
\item  one-loop \textit{virtual} corrections to $bq\rightarrow W^\pm bq^\prime$;
\item one additional parton \textit{real} emission processes:
$q\bar{q}^\prime \rightarrow W^\pm b\bar{b}+g$,
 $qg\rightarrow W^\pm b\bar{b}+q^\prime$, $bq\rightarrow W^\pm bq^\prime+g$, and $bg\rightarrow W^\pm b q \bar{q}^\prime$.
\end{itemlist}
One can notice that the subprocesses and corrections contributing to
the 4F NLO cross section for $W^\pm+1b$-jet production are a subset of
those contributing to the 5F NLO cross section, while this is not true
for $Z/\gamma+1b$ jet. In the case of $W^\pm+1b$ we can then state
that the most complete NLO QCD prediction is the 5F one, while in the
case of $Z/\gamma+1b$ jet it very much depends on the relevance of the
logarithms that are resummed in the 5F approach as compared to the
non-logarithmic terms that are neglected in the 5F but not in the 4F.

We also notice that special attention needs to be paid in
combining tree-level and higher-order corrections in order to avoid
double counting of $\Lambda_b$ logarithms that appear both in a
fixed-order process (like, e.g. $gg\rightarrow Zb\bar{b}$) from gluon
splitting, and in the corresponding $b$-initiated subprocess (like,
e.g. $bg\rightarrow Zb$) from the expansion of the $b$-quark PDF.
Double counting is avoided by introducing a \textit{subtraction term}
that removes from the the fixed-order cross section of a $b$-initiated
process the logarithmic term(s) already resummed in the $b$-quark PDF.
For example, the NLO QCD cross section for $Z+1b$ jet in the 5F should
be calculated as,
\begin{equation}
\sigma_{\mathrm{5F}}^{\mathrm{NLO}}(Z+1b)=\sigma_{bg}^{\mathrm{LO}}+\sigma_{bg}^{\mathrm{NLO}}-
\int dx_1dx_2 \tilde{b}(x_1,\mu)g(x_2,\mu)\sigma^{\mathrm{LO}}_{bg}\,\,\,,
\end{equation}
where the last term is the NLO subtraction term which avoids double
counting of the $\Lambda_b$ logarithm already included in the lowest
order $b$ PDF given in Eq.~(\ref{eq:b_pdf_lo}).

It is clear by looking at the different perturbative structure of the
calculation of $W^\pm b\bar{b}$ vs. $Z/\gamma\,b\bar{b}$ production
that we can expect 4F or 5F predictions to give a more accurate
prediction of these processes depending on the energy (Tevatron vs.
LHC) and kinematical regime. Apart from more specific situations on which
we will comment in Section~\ref{sec:results} when we discuss the
comparison of existing experimental measurements to the corresponding
theoretical predictions, we can conclude this synopsis of 4F vs. 5F by
emphasizing some general criteria. First of all, it is worth
emphasizing one more time that $V+2b$-jet processes are very well
described in the 4F at NLO in QCD. This is clear by a simple
inspection of the contributing parton-level processes and is supported
by the measurements of $W+2b$ jets (see Table~\ref{table:Wb-xs-exp-th}
and Figure~\ref{fig:Wb-XS-ATLAS}), $Z/\gamma^*+2b$ jets (see
Table~\ref{table:Zb-xs-exp-th} and Figures~\ref{fig:Zb-XS-ATLAS}
and~\ref{fig:Zb-dXS}), and $\gamma+2b$ jets (see
Figure~\ref{fig:gammabb-dXSD0}). On the other hand, for $V+1b$-jet
processes 4F and 5F approaches can be complementary. In general, a 5F
approach lends itself better to the calculation of more inclusive
quantities (like total cross sections) and, in this case, it often
allows to push the perturbative fixed order slightly higher since, in
most cases, it involves simpler processes. At the same time a 4F
approach can be preferable for the calculation of more exclusive
observables, like processes with more than a single heavy jet in the
final state.

There are however cases in which one approach can be clearly
preferable than the other beyond general considerations.  For
instance, processes that, at a given energy or in a given kinematical
regime, are determined at LO by $q\bar{q}$-initiated subprocesses
will be less sensitive to large logarithms $\Lambda_b$ that have their
origin in the integration over the low-$p_T$ region of $b$ quarks
produced in $g\rightarrow b\bar{b}$ initial-state splitting. This is
for instance the case of $W+b$ jet, where the sensitivity to
$\Lambda_b$-like logarithms is indeed a NLO effect (see results in
Table~\ref{table:Wb-xs-exp-th} and Figures~\ref{fig:Wb-XS-ATLAS}
and~\ref{figWbdXS}), or the case of $\gamma+b$-jet production at the
Tevatron, where the $q\bar{q}$-channel dominates over the $gg$-channel
production (see results in Figures~\ref{fig:gammab-dXSCDF},
\ref{fig:gammab-dXSD0} and \ref{fig:gammabb-dXSD0}).

Ideally, a 5F calculation with sufficiently high orders of QCD
corrections included would give the best of both worlds, since it
would include both enough fixed-order non-logarithmic terms (also
present in the corresponding 4F calculation) and the resummation of
several orders of potentially large initial-state logarithms. The more
perturbative orders are included the broader is the overlap between 4F
and 5F predictions for a given cross section. It is in general a good
sanity check to assure that the 4F and 5F predictions confirm the
existence of such overlap by showing more and more compatibility in
going from LO to NLO, or, if needed, to higher order.

\subsection{Theoretical uncertainties}
\label{subsec:theory_syst}

Theoretical predictions are affected by both perturbative and
non-perturbartive uncertainties. On one hand, hard-scattering cross
sections are inherently approximate since they are calculated at a
given order in perturbative QCD, while the algorithms adopted to
implement the following radiation of partons (parton showering) are
based on the approximation of accounting only for those phase-space
regions that most contribute to extra-parton radiation. On the other
hand, all hadronization models adopted to describe the formation of
hadronic bound states after the last stages of parton showering
contain some level of arbitrariness.

At the parton level, the main sources of theoretical uncertainty are
the dependence on the renormalization and factorization scales
($\mu_R$, $\mu_F$) of the NLO QCD cross sections as well as their
dependence on the choice of PDF and $\alpha_s$. The dependence on the
choice of $m_b$ can also be sizable and comparable at times to the
systematic uncertainty from PDF and $\alpha_s$. This has been
documented for instance in some $W+b$-jet
studies~\cite{Badger:2010mg,Caola:2011pz}, but is otherwise usually
neglected in experimental studies. Looking at the theoretical
predictions reported in Tables~\ref{table:Wb-xs-exp-th} and
\ref{table:Zb-xs-exp-th}, we can see that uncertainties due to scale
variation are typically at the 10-25\% level, depending on the
process, while the uncertainty from PDF and $\alpha_s$ is closer to
2-8\%.  On the other hand, from the entry in
Table~\ref{table:Wb-xs-exp-th} we see that D0 estimates the
uncertainty due to $m_b$ at the 5\% level.

It is difficult to summarize the scale dependence of different
$V+b$-jet processes in a way that could fit the various experimental
frameworks in which they are measured, since the behavior of the
theoretical predictions depends in part on the specific signatures
required by the experiments.  There are however some general
characteristics that only depend on the intrinsic QCD structure of
these processes and we will review them here.  In
Section~\ref{sec:results} we complement this information by collecting
in Tables~\ref{table:Wb-xs-exp-th} and \ref{table:Zb-xs-exp-th}
existing experimental results and the quoted theoretical predictions,
including their uncertainties, and we will add comments on the
specific of each case if necessary.

From several general studies appeared in the literature over the last
few
years~\cite{FebresCordero:2006sj,Cordero:2009kv,Badger:2010mg,FebresCordero:2008ci,Hartanto:2013aha,Campbell:2008hh,Caola:2011pz}
we can see that the cross section for all $V+2\ b$-jet processes shows
a large residual scale dependence even after full NLO QCD corrections
have been included. As explained in Section~\ref{subsec:4fns-5fns} the
calculation of $V+2\ b$ jets is a 4F-only result, and the residual
scale dependence is mainly due to the opening of a new $qg$ channel
($qg\rightarrow Vb\bar{b}+q^{(\prime)}$) at NLO. This subprocess
contributes to the $O(\alpha_s)$ real corrections of $Vb\bar{b}$
production and, as a tree level process, affects the cross section
with the large scale uncertainty of a LO process. This is most visible
in the case of $W^\pm b\bar{b}$ production, since at LO $W^\pm
b\bar{b}$ is only $q\bar{q^\prime}$- not gluon-induced. At the LHC,
where the gluon density dominates over the light-quark densities,
$qg$-initiated processes are enhanced compared to $q\bar{q}$ processes
and the impact of the $qg$-initiated $O(\alpha_s)$ corrections induce
a large scale dependence of the NLO cross section. This effect is
milder, although still very visible, for $Z/\gamma+2\ b$-jet
production, since the corresponding LO processes already contain a
$gg$-initiated channel (see Figure~\ref{fig:diags-Zbb}), and are
therefore less affected by the opening at NLO of a $qg$ channel. This
has been observed and carefully discussed in
Refs.~\refcite{FebresCordero:2006sj,Cordero:2009kv}, and~\refcite{FebresCordero:2008ci}.

On the other hand, NLO theoretical predictions for $V+1b$-jet production show in
general a more stable behavior with respect to renormalization- and
factorization-scale variations when calculated in the 5F, since the
logarithmic resummation introduced via the $b$-quark PDF also resums
scale-dependent logarithmic corrections and, as a side effect,
improves the scale dependence of the cross section. On the other hand,
the 4F predictions of $V+1b$-jet production show the same large scale
dependence as the corresponding $V+2b$-jet results. This was studied
in Refs.~\refcite{Campbell:2008hh,Caola:2011pz} for $W+1b$-jet
production and can also be seen for the case of $Z+1b$-jet production
in Table~\ref{table:Zb-xs-exp-th} where results for $Z+1b$ jet
production in both the 5F and 4F schemes are reported.

The second major source of theoretical uncertainty at parton level is
the dependence on the PDF with which the parton-level cross section is
convoluted to produce the hadronic cross section.  In the 4F the
systematic uncertainty from PDF is due to the accuracy of light-quark
and gluon parton densities. In the 5F the same is true since the
bottom PDF is defined in terms of the gluon and light-quark PDF. In
both cases, the uncertainty induced on total and differential cross
sections is of the order of 2-8$\%$ (see
Tables~\ref{table:Wb-xs-exp-th} and \ref{table:Zb-xs-exp-th}), smaller
than the uncertainty from scale dependence, and it will probably
further improve in the future. Still, the theoretical uncertainty from
PDF and $\alpha_s$ has not been systematically and consistently
implemented in all experimental studies. Many results are still
compared to theoretical predictions that use specific PDF (without
comparing to others), or do not distinguish between differences due to
the choice of $\alpha_s$ and of the PDF set.

In particular, the dependence on the $b$-quark PDF needs to be
addressed more carefully. Indeed we need to remember that the
$V+b$-jet processes can play a leading role in \textit{measuring} the
$b$-quark PDF, providing the first such direct determination and
therefore testing the range of validity of a purely perturbative
definition of the $b$-quark PDF in terms of gluon and light-quark
parton densities.  It is therefore crucial to determine the precision
with which the $b$-quark PDF itself can be extracted from $V+b$-jet
measurements.  A careful determination will have to consider all
available $V+b$-jet processes and try to select a sample of
measurements that provide enough information, when compared with
theoretical predictions, to disentangle light-quark, gluon, and
$b$-quark initiated processes.  Typically, separating samples with two
$b$ jets from samples with one $b$ jet, as well as disentangling
single $b$-jet events in which the $b$ jet contains two almost
collinear $b$ quarks or a single $b$ quark can be very helpful. To
this extent some theoretical
studies~\cite{Campbell:2008hh,Caola:2011pz} have reported predictions
for each sample separately, and recent experimental analyses, like
Ref.~\refcite{ATLAS:2012xna}, accounts for the study of techniques
developed to tag so called merged $(bb)$ jets.

Finally, it is important to mention one last source of parton-level
theoretical uncertainty, namely the effects due to double-parton
scattering,\cite{Berger:2011ep} that has been only recently formally
included in the theoretical prediction for $W^\pm+2b$ jets and
$Z/\gamma^*+2b$ jets. Double-parton scattering, where a $V+2b$
($V=W^\pm,Z/\gamma^*$) parton-level final state is produced as the
result of two scattering processes ($q\bar{q}^{(\prime)}\rightarrow V$
and $q\bar{q},gg\rightarrow b\bar{b}$), can give a substantial
contribution when the constituent subprocesses have very large cross
sections (as is the case for dijet or $b\bar{b}$ production). The
factorization on which this estimate is based, however, is a rather
crude approximation and relies on the measurement of the
effective-cross-section parameter ($\sigma_{eff}$) for hard
double-parton interactions. The large uncertainty that still affect
such measurement~\cite{Aad:2013bjm} induces a sizable error on the
corresponding $V+b$-jet cross sections (see
e.g. Table~\ref{table:Wb-xs-exp-th}).

Beyond the parton level, the best theoretical predictions can be
obtained via NLO QCD event generators (e.g. \MCNLO, \POWHEG, \SHERPA )
that interface NLO QCD parton-level calculations with parton-shower
Monte-Carlo event generators.  In this context, care must be taken to
assess and quantify the systematic uncertainties introduced by the
parton-shower algorithms and by the hadronization and underlying-event
models used.  This is a non trivial task and need to be addressed on a
process by process basis, taking into account the specifics of the
generator used and of the experimental set up. Nevertheless, general
tests aimed at estimating the dependence of the theoretical prediction
on parameters intrinsic to the showering algorithms (like scale
parameters other than renormalization and factorization scales) or the
hadronization model should be always performed and the corresponding
variations should be understood, confirmed by different approaches
(i.e. codes), and included in the theoretical systematics. Several of
the results that are collected in Tables~\ref{table:Wb-xs-exp-th} and
\ref{table:Zb-xs-exp-th} include an estimate of uncertainties from
hadronization globally quoted, together with an estimate of pile-up
and underlying-event effects, as uncertainty from
non-perturbative sources.

In this context, it should also be mentioned that implementing
$b$-quark initiated processes in NLO QCD Monte-Carlo event generators
still presents some issues. We discussed in
Section~\ref{subsec:4fns-5fns} how radiative corrections to tree-level
$b$-initiated processes includes processes involving only light quarks
in the initial state and $b$ quarks in the final state (e.g.:
$gg\rightarrow b\bar{b}\gamma$ is an $O(\alpha_s)$ real-emission
correction to $bg\rightarrow b\gamma$). In order to automatically
generate the radiative-correction processes from the corresponding
tree-level process, the $b$ quark should be consistently treated as
massless or massive in both the initial and final state. Therefore,
the traditional 5F prescription of considering a $b$ quark as massless
in $b$-initiated processes and massive when it only enters as a
final-state parton (the so called S-ACOT scheme~\cite{Kramer:2000hn})
might be impractical, if not inconsistent, in the context of a NLO QCD
event generator.  Different implementations of 5F in NLO parton-shower
Monte Carlo have considered and partially adressed the problem with
different methods, while a more rigorous solution is still being
developed.

To clarify this issue, let us observe that the treatment of processes
initiated by heavy quarks and the rigorous definition of a heavy-quark
parton density has been studied~\cite{Collins:1998rz} and there is no
conceptual obstacle in considering an initial-state heavy quark as
massive.  Indeed, a well-defined scheme to build fully-massive
heavy-quark PDF have been proposed, the ACOT
scheme~\cite{Aivazis:1993pi}, and implemented in a few sets of PDF,
namely CTEQ4-6(HQ).  However, if the heavy-quark PDF is entirely
generated via perturbative evolution of gluon and light-quark density,
it can be shown~\cite{Kramer:2000hn} that the problem is equivalent to
one in which the initial-state heavy quark is treated as massless and
its mass is retained in processes that involve heavy quarks in the
final state only. This is clearly very convenient in implementing a
5F calculation, since it allows to treat a large fraction of the
contributing subchannels and their loop corrections using a massless
heavy quark, greatly simplifying the complexity of the
calculation. The corresponding heavy-quark PDF is defined in the
S-ACOT scheme~\cite{Kramer:2000hn}, where the heavy-quark
mass ($m_b$ in our case) is kept only as a collinear regulator, and
the corresponding collinear logs are resummed using the PDF evolution
as reviewed in Section~\ref{subsec:4fns-5fns}. Slightly different
flavors of the S-ACOT scheme have been implemented in the CTEQ sets of
PDF from version 6.5 on~\cite{Pumplin:2002vw}. Other sets of PDF (MSTW,
NNPDF) implement slightly different matching schemes
(TR~\cite{Thorne:1997ga,Thorne:2006qt} for MSTW~\cite{Martin:2010db}
and FONNL~\cite{Cacciari:1998it} for NNPDF~\cite{Ball:2011mu}) to
account for the transition from above to below the heavy-quark mass
thresholds, but all rely on the calculation of the $b$ initiated
processes with $m_b=0$.

Since a consistent implementation of $b$-initiated processes in a
parton-shower NLO Monte Carlo generator naturally call for treating
the $b$ quark as massive, one should probably consider adopting the
original ACOT scheme and keeping the $b$-quark mass consistently
throughout the calculation. Provided the radiation from initial-state
massive quarks is correctly modeled in current NLO Monte-Carlo
generators, this will allow accounting for all contributing
sub-channels in each $V+b$-jet process, leaving the mutual balance
between them to naturally adapt to the different energy regimes and
kinematic regions.

Let us conclude by adding that the incremental reduction of the
current main sources of theoretical uncertainty will allow us to
consider new level of precision in the future.  For instance, there is
no estimate at the moment of the effects due to higher-order
electroweak (EW) corrections. They are expected to affect the total
cross sections by only a few percent, but they might have localized
effects on distributions, and will have to be accounted for in the
future.

\subsection{Photon isolation}
\label{subsec:ph_iso}

We would like to add in this review a special note on the issue of
photon isolation, which concerns only the $\gamma+ 1b$-jet and
$\gamma+2b$-jet processes. Since this issue is often discussed in the
corresponding experimental analyses, it is important to clarify the
difference yet compatibility between different possible theoretical
approaches. We remind the reader that this has been thoroughly
discussed in Ref.~\refcite{Hartanto:2013aha}.

Photons in a hadronic environment are usually distinguished into
\textit{prompt photons}, when they are directly produced in the hard
interaction, and \textit{secondary photons}, when they originate from
the hadronization phase of a hadronic jet or the decay of unstable
hadrons (e.g. $\pi^0\rightarrow\gamma\gamma$). While the production of
prompt photons can be described in perturbation theory, the production
of secondary photons has associated non-perturbative effects that can
only be modeled and can therefore introduce a large parametric
uncertainty in any given calculation.  Since secondary photons tend to
preeminently occur in regions of the detector with abundant hadronic
activity, in particular within or close to jets, their effect can be
eliminated by imposing so-called \textit{isolation cuts} which
specifically limit the hadronic activity around a given photon. Prompt
photons become then \textit{isolated photons} and can be easily
disentangled.

The main theoretical caveat in implementing a given prescription to
\textit{isolate} prompt photons from the hard interaction is that such
procedure can veto regions of phase space responsible for soft QCD
radiation and could therefore spoil the cancellation of infrared
divergences between virtual and real corrections in a perturbative QCD
calculation. As soon as some residual hadronic activity is admitted in
the region around the photon, very energetic collinear final-state
partons can produce a small parton-photon invariant mass and the
corresponding collinear divergences need therefore to be reabsorbed
into suitable non-perturbative fragmentation functions (FF).  The cross
section for prompt-photon production is then given by,
\begin{equation}
\sigma^\gamma (\mu_R,\mu_F,M_F) = \sigma_{\mathrm{direct}}^\gamma(\mu_R,\mu_F) + \int_0^1 dz \sum_i
\sigma_i(\mu_R,\mu_F,M_F) D_{i \rightarrow \gamma} (z,M_F),
\label{eq:photonxsec}
\end{equation}
where $\sigma_{\mathrm{direct}}^\gamma$ represents the cross section
for the direct-photon component while $\sigma_i$ denotes the cross
section for the production of a parton $i$ that further fragments into
a photon.  The fragmentation of a parton $i$ into a photon is
represented by the corresponding photon FF,
$D_{i\rightarrow \gamma}(z,M_F)$, where $z$ is the fraction of the
parton momentum that is carried by the photon, and $M_F$ is the
fragmentation scale. Examples of available FF in the literature are
by Bourhis, Fontannaz and Guillet (set I and II),\cite{Bourhis:1997yu} and by Gehrmann-de Ridder and
Glover.\cite{GehrmannDeRidder:1998ba} Fragmentation functions for
final-state partons, like PDF for
initial-state partons, are intrinsically non perturbative and
introduce into the calculation the same kind of uncertainty in the
modeling of secondary photons that one originally wanted to
eliminate. How relevant the contribution of fragmentation functions is
depends on the chosen isolation prescription.

Theoretical calculations normally use two main prescriptions denoted
as \textit{fixed-cone} and \textit{smooth-cone} (or \textit{Frixione})
prescriptions respectively. The \textit{fixed-cone} prescription is
commonly used in experiments and limits the hadronic activity inside a
cone of radius $R_0$ around the photon by imposing that the hadronic
transverse energy inside the cone does not exceed a maximum value,
$E_T^{\mathrm{max}}$, set by the experiment, i.e.
\begin{equation}
\sum_{\in R_0} E_T(\mathrm{had}) < E_T^{\mathrm{max}}\,\,\,,
\label{eq:photoiso}
\end{equation}
where $R_0 = \sqrt{\Delta \eta^2 + \Delta \phi^2}$, and $\Delta \eta$
and $\Delta\phi$ are the pseudorapidity and azimuthal-angle
differences between the photon and any hadronic activity.  After the
isolation cut, the value of $z$ is typically large, and since the FF
are dominant in the low $z$ region, the isolation procedure suppresses
the fragmentation contribution substantially.

Alternatively, the \textit{smooth-cone} isolation prescription
introduced by Frixione~\cite{Frixione:1998jh}, limits the hadronic
activity around a photon by imposing a threshold on the transverse
hadronic energy within a cone about the photon that varies with the
radial distance from the photon, i.e
\begin{equation} 
\sum_{i}  E_T^{i} \, \theta(R-R_{{i},\gamma})
 < \epsilon E_T^{\gamma}\bigg(\frac{1-\cos{R}}{1-\cos{R_0}}\bigg)^n
 \qquad \mbox{for all} \quad R \le R_0\ ,
\label{eq:frixiso}
\end{equation}
where the $i$ summation runs over all final-state partons in the
process and $E_T^{i(\gamma)}$ is the transverse energy of the parton
(photon). $R_0$ is the size of the isolation cone, $\epsilon$ is an
isolation parameter of $O(1)$, the exponent is normally set to $n=1$
or $2$, and
\begin{equation}
R_{i,\gamma}=\sqrt{(\Delta \eta_{i,\gamma})^2 + (\Delta
\phi_{i,\gamma})^2}\ . \nonumber
\end{equation}
The $\theta$-function ensures that the $i$ summation only receives
contributions from partons that lie inside the isolation cone. $R =
R_{i,\gamma}$ if there is only one parton inside the isolation cone.
The r.h.s of Eq.~\ref{eq:frixiso} vanishes as $R\rightarrow 0$, thus
the collinear configurations are suppressed while soft radiation is
allowed to be present arbitrarily close to the photon.  Since the
collinear configurations are completely removed, there is no
fragmentation component in Eq.~\ref{eq:photonxsec}.

The fragmentation contribution in the $pp(p\bar{p}) \rightarrow
b\bar{b}\gamma$ calculation is included at $O(\alpha\alpha_s^2)$. Due
to the photon-isolation requirement, a photon cannot fragment from the
tagged $b/\bar{b}$ quark.  In the $\gamma+2b$ case, the photon can
only fragment off a light parton $j$, i.e. $\sigma_i$ in
Eq.~\ref{eq:photonxsec} is the cross section for the $pp(p\bar{p})
\rightarrow b\bar{b} j$ process calculated at LO, ($\sigma_i =
\sigma_{LO}(pp(p\bar{p}) \rightarrow b\bar{b} j)$).  We notice that
$\sigma_{LO}(pp(p\bar{p}) \rightarrow b\bar{b} j)$ is finite since we
impose a cut on the photon transverse momentum. For the $1b$-tag case,
in addition to the same contribution present in the $2b$-tag case, the
photon can also fragment off an unidentified $b/\bar{b}$ quark.  The
LO $pp(p\bar{p}) \rightarrow b\bar{b} j$ cross section is divergent in
this case since the light parton in the final state can be soft and/or
collinear. To overcome this problem, one should start from the
$pp(p\bar{p}) \rightarrow b\bar{b}$ cross section at NLO in QCD.  We
also notice that when the photon is fragmented off of a $b/\bar{b}$
quark, terms proportional to $\ln(M_F^2/m_b^2)$, arising from the
collinear configuration of the $b\rightarrow b\gamma$ splitting in the
(direct) $pp(p\bar{p}) \rightarrow b\bar{b}\gamma$ process, have to be
subtracted to avoid double counting since those terms have been
included and resummed in the $b$ quark-to-photon fragmentation
function via DGLAP evolution equations.

Having implemented both methods in $\gamma+2b$ production, very little
if no difference was found in Ref.~\refcite{Hartanto:2013aha}. Similar
conclusions have been reached in other studies involving the
associated production of a photon and several
jets~\cite{Bern:2011pa,Bern:2012vx}. This should be emphasized to the
benefit of the experimental measurements, since the experimental set
up (where energy depositions are intrinsically discretized) only
allows for a \textit{fixed-cone} isolation criterion (while the
\textit{smooth-cone} criterion is a continuous prescription, even if
discrete versions can be explored).

\section{Experimental Measurements and Theory Predictions}
\label{sec:results}

Experimental measurements of vector-boson production in association
with $b$ jets started\footnote{We also note the earlier measurement in
  Ref.~\refcite{Abazov:2004zd} of the $\sigma(p\bar{p}\rightarrow Z+b
  \mathrm{jet})/\sigma(p\bar{p}\rightarrow Z+\mathrm{jet})$ ratio.} in
2008~\cite{Aaltonen:2008mt} and has been followed by more recent
studies by all the major high-energy-physics experimental
collaborations:
\begin{itemize}
 \item $W +b$ jets by CDF~\cite{Aaltonen:2009qi},
ATLAS~\cite{Aad:2011kp,Aad:2013vka}, D0~\cite{D0:2012qt,Abazov:2014fka}, and
CMS~\cite{Chatrchyan:2013uza};
\item $Z+b$ jets by CDF~\cite{Aaltonen:2008mt},
ATLAS~\cite{Aad:2011jn,Aad:2014dvb},
CMS~\cite{Chatrchyan:2012vr,Chatrchyan:2013zja,Chatrchyan:2014dha} and
D0~\cite{Abazov:2013uza};
\item $\gamma+b$ jets by D0~\cite{Abazov:2012ea,Abazov:2014hoa}, and
CDF~\cite{Aaltonen:2013coa}.
\end{itemize}
More activity is expected in the near future, since, for instance, all
the studies published so far by the LHC collaborations have been based
on $\sqrt{s}=7$~TeV data samples only. With the larger datasets
collected in Run I of the LHC at $\sqrt{s}=8$~TeV and expected in Run
II at or above $\sqrt{s}=13$~TeV, new possibilities will open and new
challenges will be met, such that more accurate measurements will
become available.

One central and common feature of all measurements mentioned is the
need for $b$-jet tagging algorithms. The main characteristic for
tagging $b$ jets is of course the presence of a secondary vertex in
the related jet. In order to increase the purity of the samples, other
requirements can be imposed on, for instance, the tagging of decay
products of the associated meson (typically leptons), or the impact
parameters, invariant mass, and number of tracks associated with the
secondary vertex.  All these variables are combined in multivariate
analyses. In the end the experiments report $b$-jet tagging
efficiencies between 35\% and 50\%, considerably lower than
$b$-tagging efficiencies around $85\%$ reported, for instance, in
studies of $t\bar{t}$ production (see for example
Ref.~\refcite{Chatrchyan:2012jua}).  Indeed one must observe that,
with the tagging algorithms employed in $V+b$-jet analyses, light-jet
mistag rates are actually very low, at the per-mille level for light
jets and at the few-percent level for $c$ jets. This is necessary when
studying $V+b$ jet signals, since much lower jet multiplicities are
explored compared to $t\bar t$ studies. Unlike $t\bar t$ production,
where one can also exploit kinematical constraints to exclude signals
from light QCD jets, $V+b$'s signatures can be overwhelmed by
backgrounds like $V+1\ {\rm light}$-jet with the light jet
misidentified as a $b$ jet. The net effect is that a high-purity 
$b$-tagging algorithm has a reduced efficiency.

An important issue in $b$-jet tagging is the possibility of
constraining the number of $B$ hadrons in the tagged jet.  Many of the
algorithms employed so far look for jets that contain a $B$ hadron, in
principle associated with a $b$ quark from the original hard
interaction. But it has been observed that care must be taken with
jets that contain pairs of $B$ hadrons which can be associated with a
gluon splitting into an almost collinear $b\bar b$ pair. These jets
are called \textit{merged $b$ jets} and denoted as $(bb)$
jets. Contributions of $(bb)$ jets in studies of signatures with $b$
jets can be seen as reducible backgrounds.  From the theoretical side,
it is known that gluon splitting into heavy quarks needs special
attention and improved modeling in parton showers (see discussion in
Section~\ref{sec:theory}).  Strategies to reduce these backgrounds are
then advantageous. In particular, studies like the one presented by
ATLAS in Ref.~\refcite{ATLAS:2012xna}, where techniques were developed
to tag $(bb)$ jets, are very important. ATLAS indeed found that they
could develop algorithms to reject $(bb)$ jets at the 95\% level,
while retaining $b$ jets with a 50\% efficiency.  We notice that this
latter study shows that a measurement on merged $b$ jets is possible,
a result that should be of great interest to the theory community that
wants to improve the description of $g\rightarrow b\bar b$ splitting
in parton showers.

In the following subsections we review the most recent experimental
measurements of $V+b$ jets ($V=W^\pm,Z/\gamma^*,\gamma$) and highlight
the comparisons to theoretical predictions that have been reported in
the corresponding experimental papers. In Section~\ref{subsec:Wb} we
present results for $W+b$-jet production, in Section~\ref{subsec:Zb}
for $Z/\gamma^*+b$-jet production, and in Section~\ref{subsec:gammab}
for $\gamma+b$-jet production. We do not present all the details of
the experimental analyses, but limit ourselves to show tables and
plots that well capture and summarize the main features of these
studies with particular emphasis on those aspects that have a direct
influence on the comparison with theoretical predictions. In
particular, Tables~\ref{table:Wb-sign-kin}, \ref{table:Zb-sign-kin},
and \ref{table:gammab-sign-kin} present a synopsis of the experimental
set up for each analysis. In the first column we give the experiment
(first the Tevatron experiments, CDF and D0, followed by the LHC
experiments, ATLAS and CMS), the center-of-mass energy, the data set
used, and the paper from where the results have been taken. In the
second column we list the signatures that have been selected by each
experiment, and in this context we denote by $b$ a $b$ jet and by $j$
a light jet, while we denote inclusive measurements (usually denoted
in the experimental papers by \textit{at least one $b$ jet} or
\textit{at least one $b$ and one light jet} or similar) adding a $+X$
to the main jet signature. Finally, in the third column we report the
detailed kinematical cuts and vetos employed at the particle level,
which should directly correspond to what has been used in obtaining
the corresponding theoretical predictions.  Experimental results and
theoretical predictions for total cross sections (and ratios of) are
listed in Tables~\ref{table:Wb-xs-exp-th}
and~\ref{table:Zb-xs-exp-th}. We have extracted the theoretical
results from the corresponding experimental papers, and therefore they
are not often directly comparable among themselves since they might
correspond to different set ups and different calculations. We will
highlight these differences in the discussion and emphasize what we
can learn from the comparison.  We will also include several figures
from different studies to illustrate the comparison between
experimental measurements and theoretical predictions of total and
differential cross sections

\subsection{Measurements of $W$ hadronic production in association with $b$ jets}
\label{subsec:Wb}

All studies presented for $W+b$-jet hadronic production consider
leptonic decays of the $W$ boson, employing either the electron
channel ($W\rightarrow e\nu_e$) or the muon channel ($W\rightarrow
\mu\nu_\mu$) or both, while the decay into a tau lepton is considered
as a background. Consequently the signatures studied contain missing
transverse energy, $\slashed{E}_{\rm T}$, due to the transverse
momentum of the escaping neutrino, $p_{\rm T}^\nu$. This
$\slashed{E}_{\rm T}$ is measured in the detectors by adding
(vectorially in the transverse plane) energy depositions around the
detector.  In theory calculations such $\slashed{E}_{\rm T}$ is
identified with $p_T^\nu$, which in experimental measurements can not
directly be measured. In what follows we choose to always write
$\slashed{E}_{\rm T}$ while quoting kinematical cuts for the different
experimental studies, as missing transverse energy is the fundamental
object that can be studied experimentally.\footnote{We do this in
  spite of the fact that all the experimental studies use $p_{\rm
    T}^\nu$ in constraining different signatures, because we think it
  is more appropriate. In fact, it would be better to identify
  $\slashed{E}_{\rm T}$ with the negative sum of the momenta of the
  observed jets (light or heavy) in the transverse plane.}

The main backgrounds that affect the $W(\rightarrow l\nu_l)+b$-jet
signatures are: $W+c$ jets and $W+$light jets where one (or more)
jet(s) is mistagged as a $b$-jet; $t\bar t$ and single-top production;
and multijet production (with some of the light jet mimicking
leptons). Also Drell-Yan processes (with extra light jets) and diboson
production can have relevant contributions as backgrounds. Kinematical
cuts and vetos are often applied to minimize background, for example
in extra leptons or jets to avoid contamination from top-pair
production.


\begin{table}[h]
\tbl{$W+b$-jet experimental signatures and kinematics from all major
high-energy-physics experiments. Details are presented in the text.}
{\begin{tabular}{@{}c|c|c@{}} \toprule
  \hline
  Experiment & Signatures & Kinematics \\
  \hline
  CDF~\cite{Aaltonen:2009qi}, $1.96$ TeV, $1.9\ {\rm fb}^{-1}$ & 
  \begin{tabular}{l}$W(\rightarrow l\nu_l)+b$\\ $W(\rightarrow l\nu_l)+b+j$ \\ $W(\rightarrow l\nu_l)+b+b$\end{tabular} & 
  \begin{tabular}{l} $p_{\rm T}^l>20$ GeV, $|\eta^l|<1.1$, \\ $\slashed{E}_{\rm T}>25$
		GeV \\ \hline {\it Jets:}\\ Cone-based algorithm, $R=0.4$\\ 
		$E_{\rm T}^j>20$ GeV, $|\eta^j|<2$ \end{tabular} \\
  \hline
  \hline
  D0~\cite{D0:2012qt}, $1.96$ TeV, $6.1\ {\rm fb}^{-1}$  & 
  $W(\rightarrow l\nu_l)+b+X$ & 
  \begin{tabular}{l} {\it Muon channel:} \\ $p_{\rm T}^\mu>20$ GeV, $|\eta^\mu|<1.7$, \\ 
		\hline {\it Electron channel:}\\ $p_{\rm T}^e>20$ GeV,\\ 
		$|\eta^e|<1.1$, or $1.5<|\eta^e|<2.5$ \\
		\hline $\slashed{E}_{\rm T}>25$ GeV \\
		\hline {\it Jets:}\\ Midpoint algorithm, $R=0.5$\\ 
		$p_{\rm T}^b>20$ GeV, $|\eta^b|<1.1$ \end{tabular}  \\
  \hline
  \hline
  ATLAS~\cite{Aad:2013vka}, $7$ TeV, $4.6\ {\rm fb}^{-1}$  & 
  \begin{tabular}{l}$W(\rightarrow l\nu_l)+b$\\ $W(\rightarrow l\nu_l)+b+j$ \\ $W(\rightarrow l\nu_l)+b+b$\end{tabular} & 
  \begin{tabular}{l} $p_{\rm T}^l>25$ GeV, $|\eta^l|<2.5$, \\ 
		$\slashed{E}_{\rm T}>25$ GeV, $M_{\rm T}^W>60$ GeV \\
		\hline {\it Jets:}\\ Anti-$k_{\rm T}$ algorithm, $R=0.4$\\ 
		$p_{\rm T}^j>25$ GeV, $|y^j|<2.1$\\ $\Delta R(l,j)>0.5$ \end{tabular}  \\
  \hline
  \hline
  CMS~\cite{Chatrchyan:2013uza}, $7$ TeV, $5.0\ {\rm fb}^{-1}$  & 
  $W(\rightarrow \mu\nu_\mu)+b+b$ & 
  \begin{tabular}{l} $p_{\rm T}^\mu>25$ GeV, $|\eta^\mu|<2.1$, \\ 
		$\slashed{E}_{\rm T}>25$ GeV \\ \hline
		{\it Jets:}\\ Anti-$k_{\rm T}$ algorithm, $R=0.5$\\ 
		$p_{\rm T}^b>25$ GeV, $|\eta^b|<2.4$\\ \hline {\it Jet Veto:}\\
		$p_{\rm T}^j>25$ GeV, $|\eta^j|<4.5$
		 \end{tabular}  \\
 \hline
\hline
\end{tabular}
\label{table:Wb-sign-kin}}
\end{table}

\subsubsection{Experimental setups and total cross sections}

In Tables~\ref{table:Wb-sign-kin} and \ref{table:Wb-xs-exp-th} we
present a detailed account of the most recent studies of each of the
major high-energy-physics experimental collaborations for
$W+b$-jets hadronic production.

Experimental details are shown in Table~\ref{table:Wb-sign-kin}.  In
the \textit{signature} column we show the lepton channels employed,
either $W(\rightarrow e\nu_e)$ or $W(\rightarrow\mu\nu_\mu)$ for
the electron or muon channel respectively, and when these are combined we
write $W(\rightarrow l\nu_l)$.  Notice that tau decays are actually
considered as background and no hadronic decay mode is considered
since they are affected by large backgrounds.  We pay special
attention to the jet multiplicity and, using the notation explained at
the beginning of this section, we distinguish between exclusive and
inclusive ($\cdots+X$) measurement and we specify how many $b$ jets
and light jets are part of the signature. This is very important in
understanding the comparison with theory.  The kinematical cuts shown
in Table~\ref{table:Wb-sign-kin} define the phase space considered to
produce theoretical predictions at the particle level. We note that
experimental analyses use more sophisticated selections, tailored to
the optimization of the experimental apparatus and the available data,
and subsequently \textit{transform} these, via an \textit{unfolding}
procedure, to a set of simpler cuts and vetos to be used in theory
Monte Carlo generators. Care must be taken to reduce as much as
possible the unfolding procedure in order to minimize the sensitivity
to the modeling of data in the Monte Carlo employed by the experiment.


\begin{table}[h]
  \tbl{$W+b$-jet measurements by the major
    high-energy-physics experiments and corresponding theoretical
      predictions. 
    The CDF measurement corresponds to a jet cross
    section, while all others are event cross sections. Details are presented in the text.}
  {\begin{tabular}{@{}c|l@{}}\toprule
      \hline
      Setup & Cross sections in pb \\
      \hline
      CDF~\cite{Aaltonen:2009qi}  & 
  \begin{tabular}{lcl} Experiment & $2.74$ & $\pm 0.27({\rm stat.})\pm 0.42({\rm syst.})$ \\ 
		Theory & $1.22$ & $\pm 0.14({\rm scale})$
  \end{tabular} \\
  \hline
  \hline
  D0~\cite{D0:2012qt}  & 
  \begin{tabular}{lcl} Experiment ({\it Muon channel}) & $1.04$ & $\pm 0.05({\rm stat.})\pm 0.12({\rm syst.})$ \\ 
		Theory ({\it Muon channel}) & $1.34$ & $^{+0.40}_{-0.33}({\rm scale}) \pm
			0.06({\rm PDF}) ^{+0.09}_{-0.05} (m_b)$ \\ \hline
		Experiment ({\it Electron channel}) & $1.00$ & $\pm 0.04({\rm stat.})\pm 0.12({\rm syst.})$ \\ 
		Theory ({\it Electron channel}) & $1.28$ & $^{+0.40}_{-0.33}({\rm
			scale}) \pm 0.06({\rm PDF}) ^{+0.09}_{-0.05} (m_b)$ \\ \hline
		Experiment ({\it Combined}) & $1.05$ & $\pm 0.03({\rm stat.})\pm 0.12({\rm syst.})$ \\ 
		Theory ({\it Combined}) & $1.34$ & $^{+0.40}_{-0.33}({\rm scale}) \pm
			0.06({\rm PDF}) ^{+0.09}_{-0.05} (m_b)$
  \end{tabular} \\
  \hline
  \hline
  ATLAS~\cite{Aad:2013vka}  & 
  \begin{tabular}{lcl} Experiment (1 jet) & $5.0$ & $\pm 0.5({\rm stat.})\pm 1.2({\rm syst.})$ \\ 
		Theory (1 jet) & $3.01$ & 
			\begin{tabular}{l} $\pm 0.07({\rm stat.})
				^{+0.72}_{-0.54}({\rm scale})\pm 0.04({\rm PDF})$ \\ 
				$\pm 0.08({\rm NP}) ^{+0.40}_{-0.29} ({\rm DPI})$ 
			\end{tabular} \\ \hline
		Experiment (2 jets) & $2.2$ & $\pm 0.2({\rm stat.})\pm 0.5({\rm syst.})$ \\ 
		Theory (2 jets) & $1.69$ & 
			\begin{tabular}{l} $\pm 0.06({\rm stat.})
				^{+0.40}_{-0.23}({\rm scale})\pm 0.04({\rm PDF})$ \\ 
				$\pm 0.08({\rm NP}) ^{+0.12}_{-0.09} ({\rm DPI})$ 
			\end{tabular} \\ \hline
		Experiment (1 or 2 jets) & $7.1$ & $\pm 0.5({\rm stat.})\pm 1.4({\rm syst.})$ \\ 
		Theory (1 or 2 jets) & $4.70$ & 
			\begin{tabular}{l} $\pm 0.09({\rm stat.})
				^{+0.60}_{-0.49}({\rm scale})\pm 0.06({\rm PDF})$ \\ 
				$\pm 0.16({\rm NP}) ^{+0.52}_{-0.38} ({\rm DPI})$ 
			\end{tabular}
  \end{tabular} \\
  \hline
 \hline
  CMS~\cite{Chatrchyan:2013uza}  & 
  \begin{tabular}{lcl} Experiment & $0.53$ & $\pm 0.05({\rm stat.})\pm 0.09({\rm
			syst.})\pm 0.06({\rm theo.}) \pm 0.01({\rm lumi.})$ \\ 
		Theory & $0.55$ & $\pm 0.03({\rm scale \& PDF})\pm 0.01({\rm
			NP})\pm 0.05({\rm DPI})$
  \end{tabular} \\
  \hline
  \hline
\end{tabular}
\label{table:Wb-xs-exp-th}}
\end{table}

The results of the measurements introduced in
Table~\ref{table:Wb-sign-kin} are shown in
Table~\ref{table:Wb-xs-exp-th}, where we also collect some of the
theoretical predictions quoted by the experimental papers. CDF reports
jet cross sections, while all other experiments give event cross
sections.  We present the full set of reported uncertainties. For
experimental results these are of statistical and systematic
nature. Also, CMS separate the uncertainty due to Monte Carlo modeling
in the extraction of selection efficiencies (``theo.")  and the one
due to luminosity determination.  Experimental statistical
uncertainties are relatively low, of the order of 10\%, for all
measurements.
It is to be expected that statistical errors at the
total-cross-section level will be relatively marginal for all future
LHC studies (either at $\sqrt{s}=8$ TeV or at $\sqrt{s}=13$ TeV). The
challenge for the experiments will then be to reduce systematic
uncertainties (which so far vary between 10\% and 30\%) through a
better understanding of their detectors. In particular, reducing the
systematic uncertainty associated to $b$-jet tagging procedures might
be crucial. For ATLAS and CMS, the larger luminosity environment of
RUN II, with the larger number of pile-up events, will pose serious
difficulties.

The theory predictions listed in Table~\ref{table:Wb-xs-exp-th} are
all parton-level predictions that include NLO QCD corrections as
implemented in MCFM following
Refs.~\refcite{FebresCordero:2006sj,Campbell:2008hh,Badger:2010mg,Caola:2011pz}.
In the ATLAS and CMS studies, they have been corrected by
non-perturbative effects.  The theoretical uncertainty is separated
in: dependence on the (unphysical) factorization and renormalization
scales (``scale"), dependence on the choice of parton distribution
functions (``PDF"), dependence on the value of $m_b$, non-perturbative
corrections (``NP"), and dependence on the modeling of contributions
from double-parton interactions (``DPI").

All theory results show a relatively large unphysical scale
sensitivity around 15\% to 30\%. Although a full calculation of
Next-to-Next-to-Leading-Order (NNLO) QCD corrections might be far in
the future, it might be possible to compute the gauge-invariant pieces
most sensitive to scale variations. It has been noticed that this
sensitivity comes mainly from the tree-like NLO QCD real corrections
associated to a quark-gluon initial state. A step towards this has
been achieved by computing 1-loop amplitudes for $Wb{\bar
  b}j$~\cite{Reina:2011mb}, and even more in
Ref.~\refcite{Luisoni:2015mpa} where full NLO QCD corrections have
been computed.  We notice that CMS, which has presented a measurement
for the exclusive $W+b+b$ process, has estimated the scale sensitivity
following the prescription of Ref.~\refcite{Stewart:2011cf}. This is a
way to give a more meaningful theoretical scale uncertainty since, for
exclusive processes, the dependence on the scale variation tends to be
relatively small (given the reduced impact from tree-like real
contributions~\cite{FebresCordero:2006sj,Cordero:2009kv}).  PDF
uncertainties are mild and of the order of 2\% to 4\%.
Non-perturbative uncertainties are mainly associated to hadronization
and underlying-event models.  ATLAS calculates them using the
implementation of $Wb\bar{b}$ production in \POWHEGBOX, which however
is a purely 4F calculation and might not give the correct information
for $W+b+X$ (while it would for $W+b+b$ or $W+b+b+X$). CMS, on the
other hand, estimates hadronization corrections comparing with a
tree-level simulation obtained using \MadGraph+\PYTHIA. Since CMS
measured a $W+b+b$ cross section, using the aforementioned
implementation in \POWHEGBOX{} or in any other NLO parton-shower Monte
Carlo generator (as mentioned in the introduction, $Wb\bar{b}$ is also
available e.g. in \MadaMCNLO) would have provided a more adequate
information.  In future predictions we cannot but recommend the
systematic use of NLO QCD parton-shower Monte Carlo generators to
directly assess the impact of these non-perturbative effects.  Finally
it is important to mention the large theory uncertainty quoted by
ATLAS and CMS as due to double-parton interactions (DPI).  ATLAS
estimates that double-parton interaction has a large impact, of the
order of 25\%, on the total cross section, while CMS accounts for an
effect at the 15-10\% level.  Even more, ATLAS shows that, when quoted
differentially in $b$-jet $p_{\rm T}$, these contribution mostly fall
in the lower bins. It is interesting to notice that, as we will see in
Section~\ref{subsec:Zb}, the contribution of DPI to the
$Z/\gamma^*+b$-jet cross section is estimated by CMS to be below
5\%~\cite{Chatrchyan:2014dha}, also occurring mainly in the lower bins
in $p_{\rm T}^b$. This difference might be due to the better
kinematical resolution achieved via the decaying products of the
$Z/\gamma^*$ boson ($Z/\gamma^*\rightarrow l\bar{l}$ as opposed to
$W\rightarrow l\nu_l$) which allows a better modeling of DPI.  Given
the theoretical challenge involved in rigorously describing DPI in
QCD, more dedicated (data driven) studies are needed, in order to be
able to reduce the DPI impact on $W+b$-jet measurements.

\begin{figure}[ht]
\begin{center}
\includegraphics[scale=0.4]{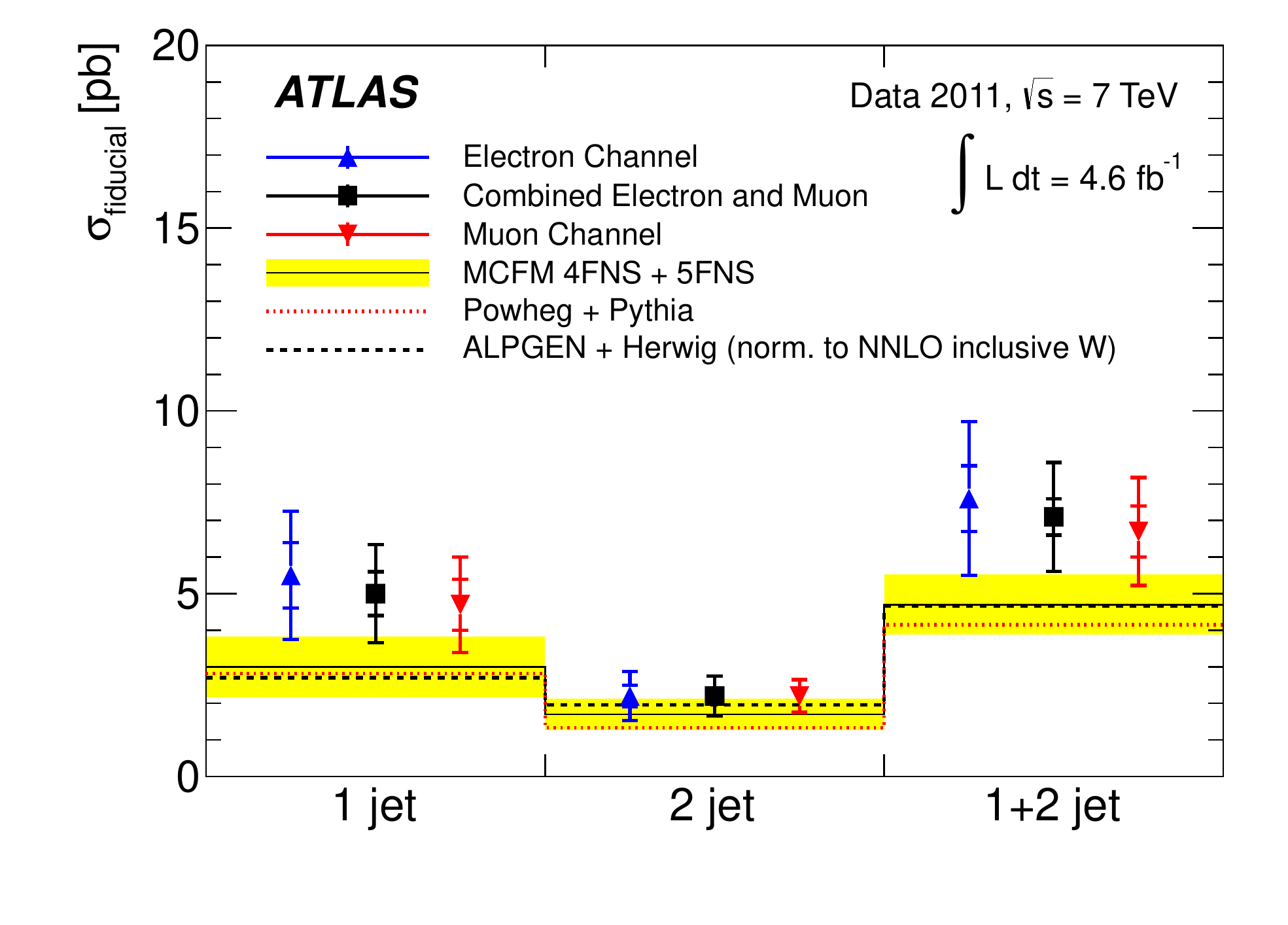}
\caption{Total cross sections for $W+b$-jet hadronic production
  measured by ATLAS~\cite{Aad:2013vka} and compared to theoretical
  predictions.}
\label{fig:Wb-XS-ATLAS}
\end{center}
\end{figure}

Finally in Figure~\ref{fig:Wb-XS-ATLAS} we present a plot with a full
account of the ATLAS measurements, including, on top of the NLO QCD
predictions of Ref.~\refcite{Caola:2011pz} (MCFM 4FNS+5FNS) other
Monte-Carlo-based theory predictions: POWHEG+Pythia, based on the 4F
calculation of $Wb\bar{b}$ but matched to a parton shower at NLO in
QCD~\cite{Oleari:2011ey}, and ALPGEN+Herwig, matched to parton shower
but without full NLO QCD corrections. Also, results are reported in
jet-multiplicity bins and offer the possibility to assess separately
the comparison between theory and experiments in each case.

Looking at the results in Table~\ref{table:Wb-xs-exp-th} and
Figure~\ref{fig:Wb-XS-ATLAS}, one can see that the agreement between
experiments and theory is relatively good, although within still
sizable uncertainties. We notice that the agreement is definitely
better for signatures with at least two jets. For single-$b$-jet
signatures, CDF does have a large excess with significance around
$3\sigma$. On the other hand, D0 does not see such discrepancy, but it
is important to notice that the measurement by D0 is fully inclusive
and it is an event cross section. Also, given the difference in their
(both IR unsafe) jet algorithms, the comparison is difficult. On their
side, ATLAS measurements are higher than the corresponding theory
predictions, but with little significance (around $1.5\sigma$). It
will be interesting to see if such tension persists when ATLAS
publishes updates with larger data sets. The comparison between
different theoretical predictions given in
Figure~\ref{fig:Wb-XS-ATLAS}, with and without parton-shower and
non-perturbative effects, shows clearly that different theoretical
estimates are in relatively good agreement, giving confidence in
particular in the quoted NLO uncertainty bands. Finally, we note that
the slight experimental excess is shown clearly to come from the 1-jet
bin. This seems to be confirmed by the complementary CMS measurement
which focus on the $W+b+b$ signature, and find excellent agreement
between theory and experiment.
\begin{figure}[ht]
\includegraphics[scale=0.3]{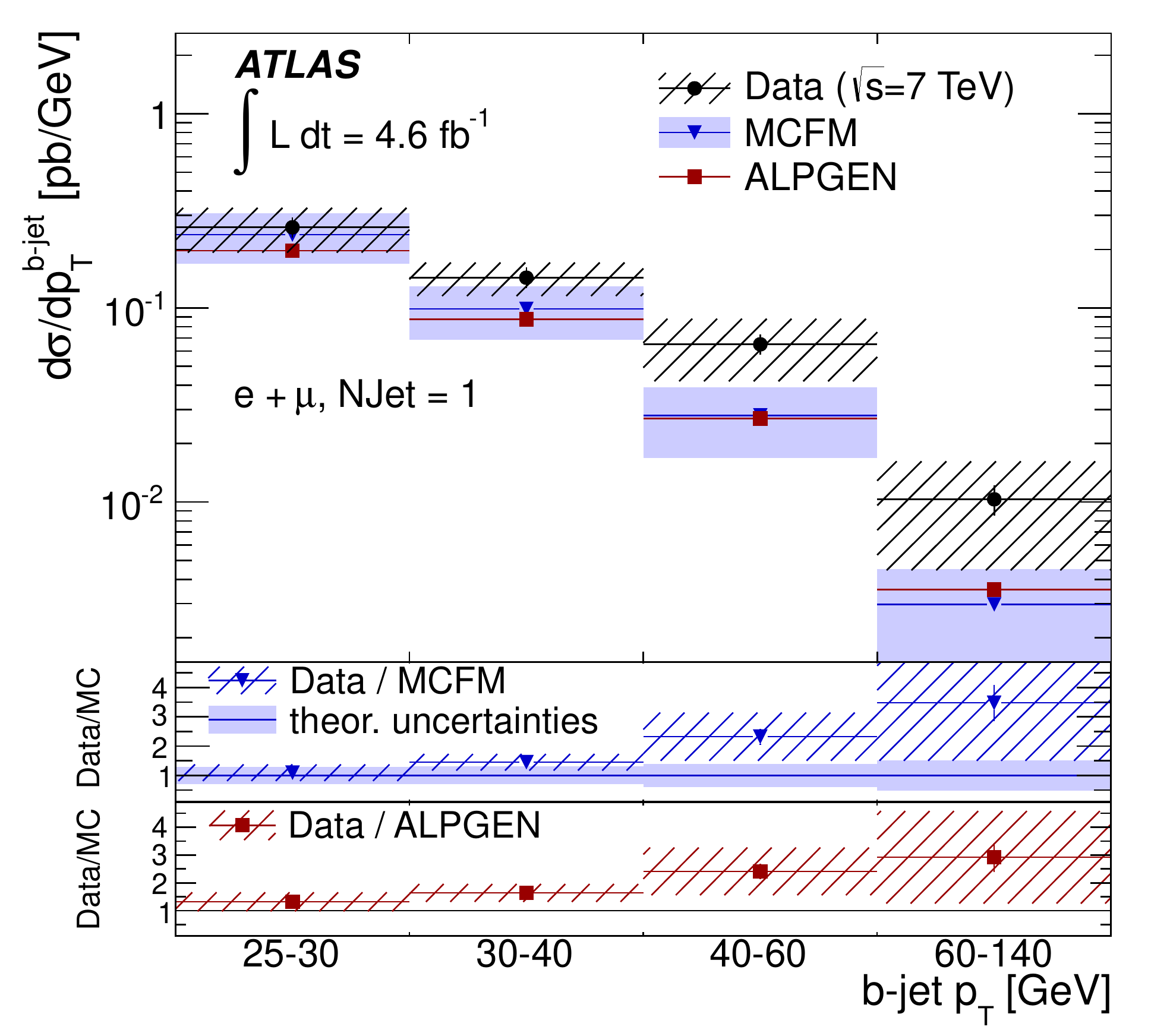}
\hspace{2mm}
\includegraphics[scale=0.31]{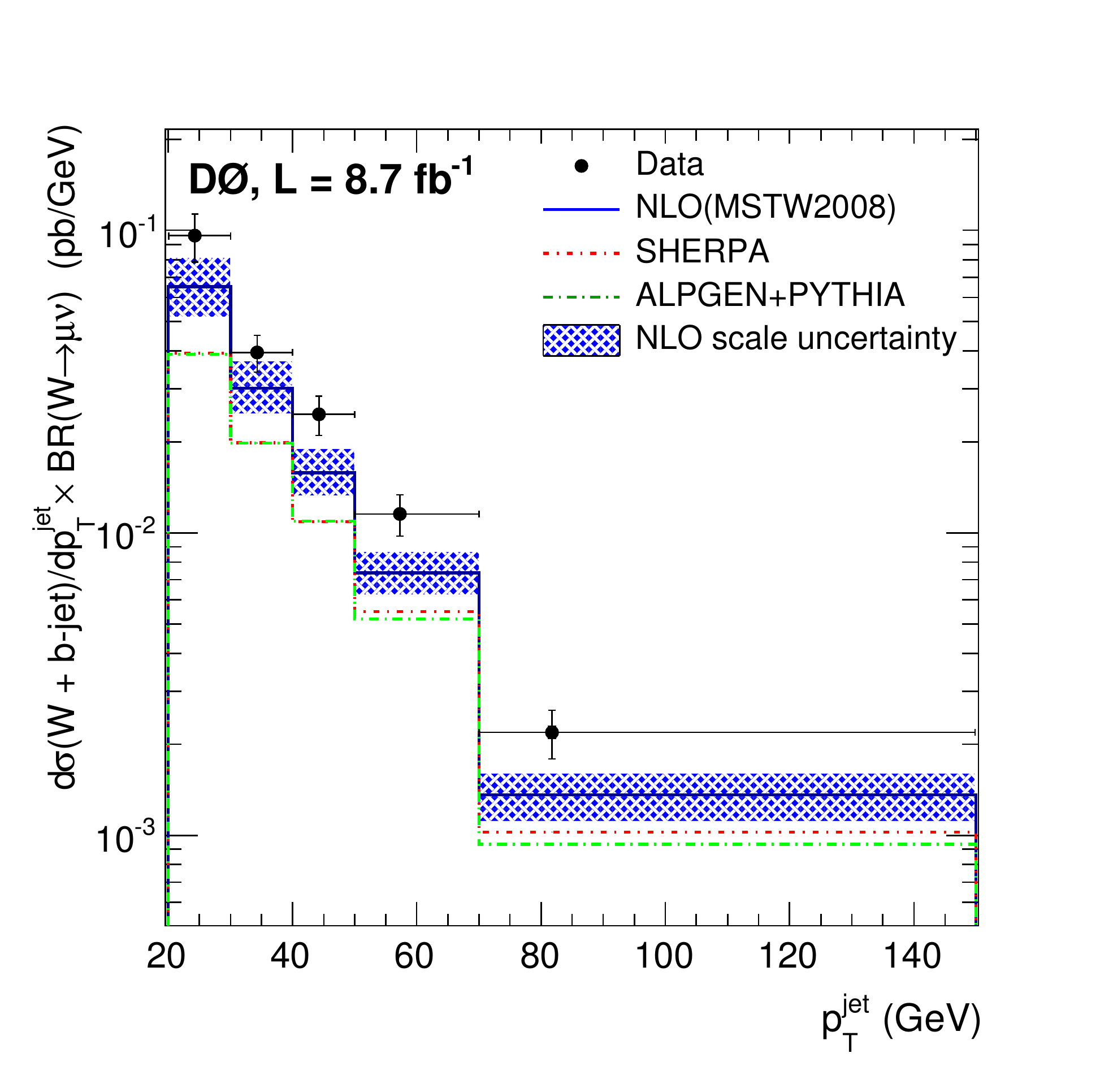}
\caption{Differential cross sections for various $W+b$-jet
  signatures, as a function of the $b$-jet $p_{\rm T}$, measured by
  ATLAS~\cite{Aad:2013vka} (l.h.s.) and D0~\cite{Abazov:2014fka}
  (r.h.s.), and compared to theoretical predictions.}
\label{figWbdXS}
\end{figure}

\subsubsection{Differential cross sections}

In order to show how well experiments and theory compare over phase
space, we show in Figure~\ref{figWbdXS} a couple of plots from
ATLAS~\cite{Aad:2013vka} and D0~\cite{Abazov:2014fka}. Notice that the
differential cross section from D0 comes from a more recent study as
compared to the one quoted in Table~\ref{table:Wb-sign-kin}.

Looking at the ATLAS differential cross section as a function of
$b$-jet $p_{\rm T}$ it seems that the theory predictions deviate
considerably from data at large transverse momentum, where, however,
uncertainties are also bigger. On the other hand, the D0 spectrum
seems systematically higher with respect to the theoretical
predictions, although the ratio between theory and data seems
constant.  It will be interesting to study these differential
distributions more in detail with future larger data sets, and
assess the impact of DPI in various $p_{\rm T}^b$ regions, as first
studied by ATLAS~\cite{Aad:2013vka}, as well as the possibility that
large $p_{\rm T}^b$ bins might be ``polluted" by $(bb)$ jets,
i.e. jets that contain a pair of $B$ mesons.

\subsection{Measurements of $Z/\gamma^*$ hadronic
  production in association with $b$ jets}
\label{subsec:Zb}

Measurements presented for $Z/\gamma^*+b$-jet production have employed
both electron and muon channels for the decay of the intermediate
neutral gauge boson ($Z/\gamma^*\rightarrow
e^+e^-,\mu^+\mu^-$). Therefore the main signal consists of a pair of
opposite-sign leptons and associated $b$ and light jets.\footnote{This
  signature can be generated by both $Z$ and $\gamma^*$ although the
  first component is largely dominant.} The main backgrounds in these
studies are represented by top-quark pair production, $W$ production
with jets, $Z$ production with light jets, inclusive $Z$ production
with $Z$ decaying into tau leptons, diboson and single-top
production. In order to enhance the signal, experimentalists choose
kinematical cuts and apply vetos as needed.


\begin{table}[h]
\tbl{$Z/\gamma^*+b$-jet experimental signatures and kinematics from all major
high-energy-physics experiments. Details are presented in the text.}
{\begin{tabular}{@{}c|c|c@{}}\toprule
  \hline
  Experiment & Signatures & Kinematics \\
  \hline
  CDF~\cite{Aaltonen:2008mt}, $1.96$ TeV, $2\ {\rm fb}^{-1}$  & 
  $Z/\gamma^*(\rightarrow l{\bar l})+b+X$ & 
  \begin{tabular}{l} $76\ {\rm GeV}<M_{ll}<106$ GeV \\ \hline {\it Jets:}\\ Cone-based algorithm, $R=0.7$\\ 
		$E_{\rm T}^b>20$ GeV, $|\eta^b|<1.5$ \end{tabular}  \\
  \hline
  \hline
  D0~\cite{Abazov:2013uza}, $1.96$ TeV, $9.7\ {\rm fb}^{-1}$  & 
  $\left(\frac{Z/\gamma^*(\rightarrow l{\bar l})+b+X}{Z/\gamma^*(\rightarrow
		l{\bar l})+j+X}\right)$ & 
  \begin{tabular}{l}
		$70\ {\rm GeV}<M_{ll}<110$ GeV \\
		\hline {\it Muon channel:} \\ $p_{\rm T}^{1st\ \mu}>15$ GeV,
		$p_{\rm T}^{2nd\ \mu}>10$ GeV,\\ $|\eta^\mu|<2$ \\ 
		\hline {\it Electron channel:}\\ $p_{\rm T}^e>15$ GeV,\\ 
		$|\eta^e|<1.1$, or $1.5<|\eta^e|<2.5$, \\ 
		\hline {\it Jets:}\\ Midpoint algorithm, $R=0.5$\\ 
		$p_{\rm T}^b>20$ GeV, $|\eta^b|<2.5$ \end{tabular}  \\
  \hline
  \hline
  ATLAS~\cite{Aad:2014dvb}, $7$ TeV, $4.6\ {\rm fb}^{-1}$  & 
  \begin{tabular}{l}$Z/\gamma^*(\rightarrow l{\bar l})+b+X$\\
			$Z/\gamma^*(\rightarrow l{\bar l})+b+b+X$\end{tabular} & 
  \begin{tabular}{l} $p_{\rm T}^l>20$ GeV, $|\eta^l|<2.5$, \\ 
		$76\ {\rm GeV}<M_{ll}<106$ GeV \\
		\hline {\it Jets:}\\ Anti-$k_{\rm T}$ algorithm, $R=0.4$\\ 
		$p_{\rm T}^j>20$ GeV, $|y^j|<2.4$\\ $\Delta R(l,j)>0.5$ \end{tabular}  \\
  \hline
  \hline
  CMS~\cite{Chatrchyan:2014dha}, $7$ TeV, $5.0\ {\rm fb}^{-1}$  & 
  \begin{tabular}{l} $Z/\gamma^*(\rightarrow l{\bar l})+b$\\ $Z/\gamma^*(\rightarrow l{\bar
		l})+b+X$\\ $Z/\gamma^*(\rightarrow l{\bar l})+b+b+X$\end{tabular} & 
  \begin{tabular}{l} $p_{\rm T}^l>20$ GeV, $|\eta^l|<2.4$, \\ 
		$76\ {\rm GeV}<M_{ll}<106$ GeV \\
		\hline {\it Jets:}\\ Anti-$k_{\rm T}$ algorithm, $R=0.5$\\ 
		$p_{\rm T}^j>25$ GeV, $|\eta^j|<2.1$\\ $\Delta R(l,j)>0.5$ \end{tabular}  \\
  \hline
\hline
\end{tabular}
\label{table:Zb-sign-kin}}
\end{table}

\subsubsection{Experimental setups and total cross sections}

In Table~\ref{table:Zb-sign-kin} we present details of the
experimental setups employed on $Z+b$-jet production studies, using
the same format of Table~\ref{table:Wb-sign-kin}.  Unlike the case of
$W+b$-jet measurements, most of the results shown here are inclusive
in the number of jets. Also, as it is needed in order to avoid the
massless photon pole, a cut on the invariant mass of the lepton pair
is always imposed around the $Z$ peak.


\begin{table}[h]
  \tbl{$Z/\gamma^*+b$-jet measurements by the major
    high-energy-physics experiments and corresponding theoretical
      predictions. 
    CDF and D0 measurements are ratio of 
    cross sections. Details are presented in the text.}
  {\begin{tabular}{@{}c|l@{}}\toprule
      \hline
      Setup & Cross sections or ratios \\
      \hline
      CDF~\cite{Aaltonen:2008mt}  & 
  \begin{tabular}{ll} Experiment (ratio to inclusive $Z$) & $(3.32$ $\pm 0.53({\rm
		stat.})\pm 0.42({\rm syst.}))\times 10^{-3}$ \\ 
		Theory (ratio to inclusive $Z$) & $(2.3\ - \ 2.8)({\rm scale})\times 10^{-3}$
  \end{tabular} \\
  \hline
  \hline
  D0~\cite{Abazov:2013uza}  & 
  \begin{tabular}{ll} Experiment (ratio) & $0.0196$ $\pm 0.0012({\rm
		stat.})\pm 0.0013({\rm syst.})$ \\ 
		Theory (ratio) & $0.0206$ $^{+0.0022}_{-0.0013}({\rm scale\& NP})$
  \end{tabular} \\
  \hline
  \hline
  ATLAS~\cite{Aad:2014dvb}  & 
  \begin{tabular}{lcl} Experiment [fb] (at least 1 $b$ jet) & $4820$ & $\pm 60({\rm stat.})\ 
		^{+360}_{-380}({\rm syst.})$ \\ 
		Theory, MCFM [fb] (at least 1 $b$ jet) & $5230$ &
			\begin{tabular}{l} $\pm 30({\rm stat.})\ ^{+690}_{-710}({\rm scale\& NP})$\\ 
				$\pm 6\%$(PDF)\end{tabular} \\
		Theory, 4FNS aMC@NLO [fb] (at least 1 $b$ jet) & $3390$ & $\pm 20({\rm stat.})\ 
		^{+580}_{-480}({\rm scale})$ \\
		Theory, 5FNS aMC@NLO [fb] (at least 1 $b$ jet) & $4680$ & $\pm 40({\rm stat.})\ 
		^{+550}_{-580}({\rm scale})$ \\
		\hline Experiment [fb] (at least 2 $b$ jets) & $520$ & $\pm 20({\rm stat.})\  
		^{+74}_{-72}({\rm syst.})$ \\ 
		Theory, MCFM [fb] (at least 2 $b$ jets) & $410$ & 
			\begin{tabular}{l} $\pm 10({\rm stat.})\ ^{+60}_{-60}({\rm scale\& NP})$ \\
				$\pm 5\%$(PDF)\end{tabular} \\
		Theory, 4FNS aMC@NLO [fb] (at least 2 $b$ jets) & $485$ & $\pm 7({\rm stat.})\ 
		^{+80}_{-70}({\rm scale})$ \\
		Theory, 5FNS aMC@NLO [fb] (at least 2 $b$ jets) & $314$ & $\pm 9({\rm stat.})\ 
		^{+30}_{-30}({\rm scale})$ \\
  \end{tabular} \\
  \hline
  \hline
  CMS~\cite{Chatrchyan:2014dha}  & 
  \begin{tabular}{lcl} 	Experiment [pb] (1 $b$ jet) & $3.52$ & $\pm 0.02({\rm stat.})\  
		\pm 0.20({\rm syst.})$ \\ 
		Theory, MCFM [pb] (1 $b$ jet) & $3.03$ & $^{+0.30}_{-0.36}({\rm scale})$ \\
		Theory, 4FNS aMC@NLO [pb] (1 $b$ jet) & $2.36$ &
			$^{+0.47}_{-0.37}({\rm scale})$ \\
		Theory, 5FNS aMC@NLO [pb] (1 $b$ jet) & $3.70$ &
			$^{+0.23}_{-0.26}({\rm scale})$ \\
 		\hline
		Experiment [pb] (at least 1 $b$ jet) & $3.88$ & $\pm 0.02({\rm stat.})\ 
		\pm 0.22({\rm syst.})$ \\ 
		Theory, MCFM [pb] (at least 1 $b$ jet) & $3.23$ & $^{+0.34}_{-0.40}({\rm scale})$ \\
		Theory, 4FNS aMC@NLO [pb] (at least 1 $b$ jet) & $2.71$ &
			$^{+0.52}_{-0.41}({\rm scale})$ \\
		Theory, 5FNS aMC@NLO [pb] (at least 1 $b$ jet) & $3.99$ &
			$^{+0.25}_{-0.29}({\rm scale})$ \\
 		\hline
		Experiment [pb] (at least 2 $b$ jets) & $0.36$ & $\pm 0.01({\rm stat.})\  
		\pm 0.07({\rm syst.})$ \\ 
		Theory, MCFM [pb] (at least 2 $b$ jets) & $0.29$ & $^{+0.04}_{-0.04}({\rm scale})$ \\
		Theory, 4FNS aMC@NLO [pb] (at least 2 $b$ jets) & $0.35$ &
			$^{+0.08}_{-0.06}({\rm scale})$ \\
		Theory, 5FNS aMC@NLO [pb] (at least 2 $b$ jets) & $0.29$ &
			$^{+0.04}_{-0.04}({\rm scale})$ \\
  \end{tabular} \\
  \hline
  \hline
\end{tabular}
\label{table:Zb-xs-exp-th}}
\end{table}

In Table~\ref{table:Zb-xs-exp-th} we show the results of the
experimental measurements, together with the theoretical predictions
quoted by the experimental studies. Notice that, although
CDF~\cite{Aaltonen:2008mt} showed results for jet-level cross
sections, we show only results for the ratio with respect to inclusive
$Z$-boson production, since this is the observable that they compare
to theoretical predictions.  Also, as shown in
Table~\ref{table:Zb-sign-kin}, D0 measurement is a ratio of event
cross sections for inclusive production of $Z+b+X$ and $Z+j+X$.  ATLAS
and CMS results are event cross sections.  For all measurements we
present theoretical results based on parton-level calculations that
include NLO QCD results. Also, for ATLAS and CMS we include fully
showered results including NLO QCD corrections for the hard
interaction, both in the 4F and in the 5F schemes.

Reported errors are of statistical and systematic nature for
experimental measurements. For theory, reported errors correspond to
statistical integration errors (``stat."), sensitivity to unphysical
renormalization and factorization scales (``scale"), non-perturbative
corrections (``NP"), and choice of parton distribution functions
(``PDF"). The latter has been estimated only for the ATLAS results, as
they report theoretical predictions using three different sets of PDF
(MSTW2008,\cite{Martin:2009iq} CT10,\cite{Lai:2010vv} and
NNPDF23~\cite{Ball:2012cx}).

The statistical errors of the experimental measurements are all below
6\% except for the CDF result, which happens to use a fraction of
their final data set. All systematic errors are relatively small,
below 14\% (except for the inclusive two $b$-jet measurement from CMS,
which has 19\%). Theoretical uncertainties tend to be dominated by a
scale sensitivity of the order of 15\%, with PDF uncertainties
relatively low at the 5\% level. ATLAS has included non-perturbative
corrections to the parton level NLO QCD results due to DPI, and
estimated them to be of order 1\%, much smaller than the corresponding
corrections in $W+b$ jet production. As we commented in
Section~\ref{subsec:Wb}, this may be due to the better resolved
kinematics that allows a better and more constrained modeling of
$Z/\gamma^*+b$-jets DPI contributions.

\begin{figure}[ht]
\begin{center}
\hspace{-0.8truecm}
\includegraphics[scale=0.32]{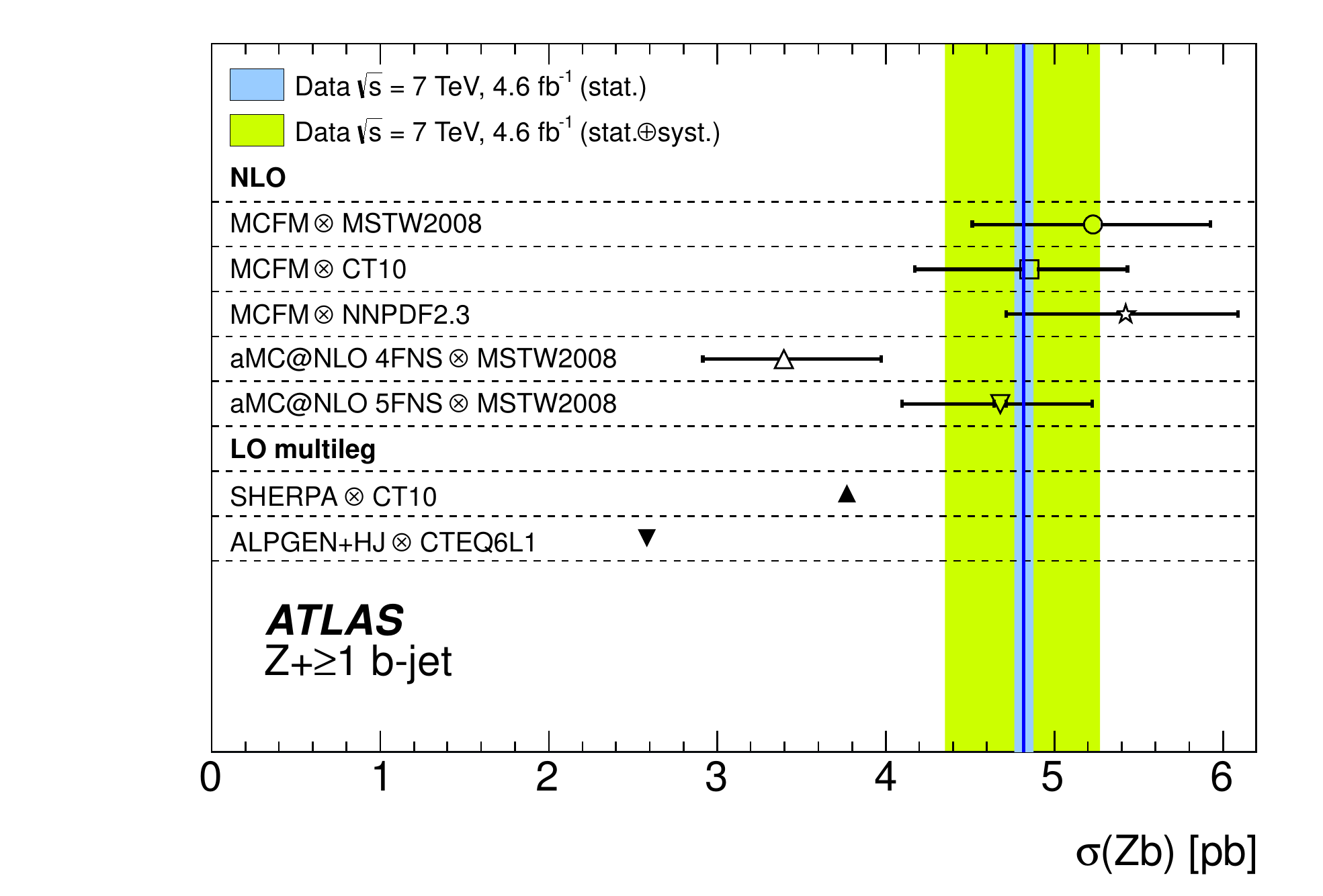}
\includegraphics[scale=0.32]{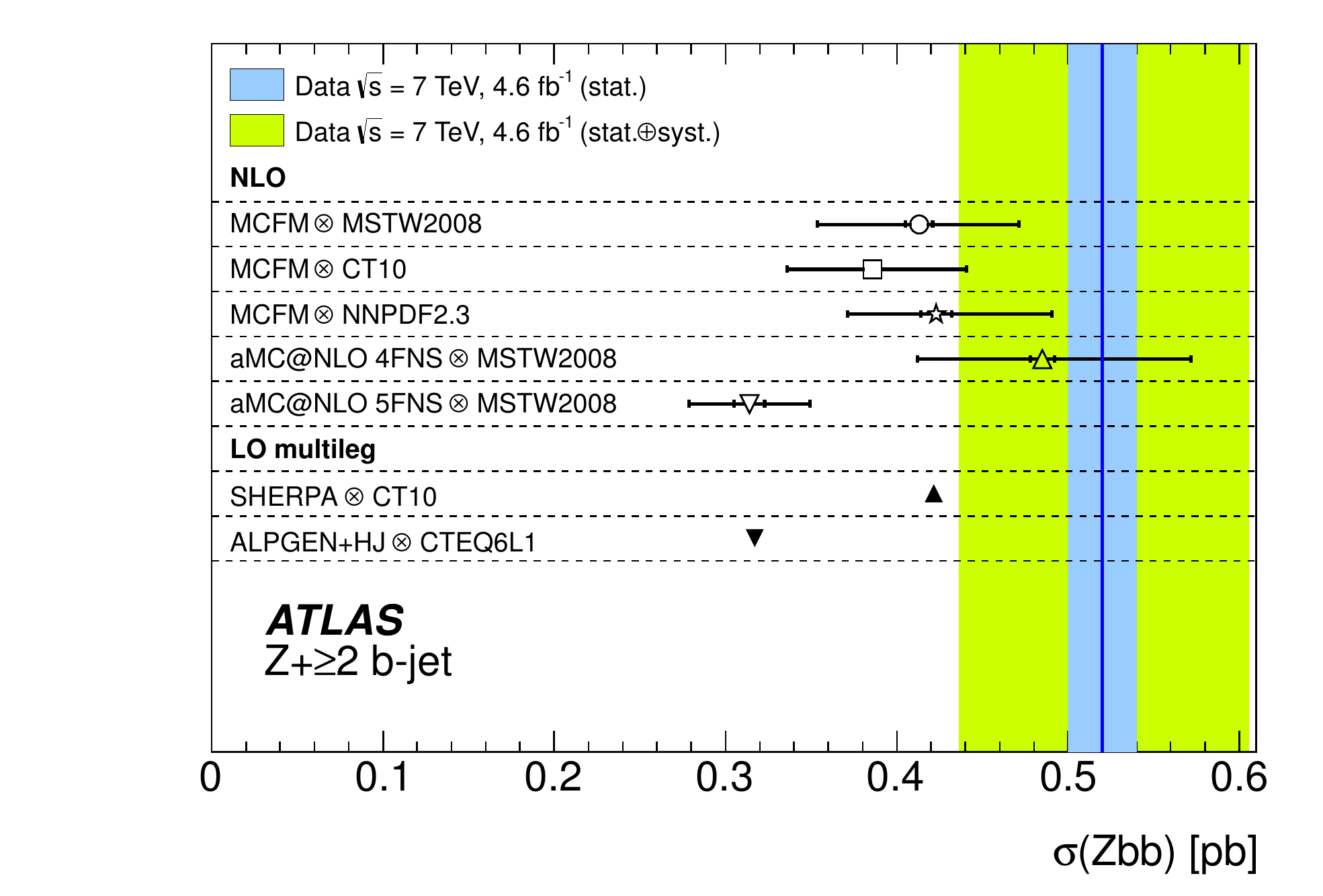}
\end{center}
\caption{Total cross sections for hadronic production of
  $Z/\gamma^*+\ge 1b$~jet and $Z/\gamma^*+\ge 2b$~jets from
  ATLAS~\cite{Aad:2014dvb} compared to various theoretical
  predictions. See text for more details.}
\label{fig:Zb-XS-ATLAS}
\end{figure}

Both ATLAS and CMS have included comparisons to theoretical results
based on parton showers that consistently include NLO QCD corrections
for the hard interaction. A summary at a glance of the ATLAS study for
both $Z/\gamma^*+b+X$ and $Z/\gamma^*+2b+X$ signatures is nicely
illustrated in Figure~\ref{fig:Zb-XS-ATLAS}.  This is an important
step, as we hope in the future to be able to consistently study all
observables including quantum corrections fully showered and
hadronized to reach hadron level predictions. Even more, when
analyzing signatures with multiple jet bins, techniques for merging
NLO QCD matrix elements for different multiplicities can be
employed~\cite{Hoeche:2014lxa}.  Doing this one has a proper way of
comparing theoretical results (with well estimated uncertainties)
directly with the observables measured by the experiments.  In
particular, both ATLAS and CMS have employed the aMC@NLO framework. In
the future, it will be relevant to systematically compare predictions
using other NLO QCD parton-shower Monte Carlo, such as the
\POWHEGBOX{} and \SHERPA.

Comparing experimental measurements to the corresponding theoretical
results obtained at NLO QCD or adding parton shower, in the 4F and 5F
schemes, one can notice that shower and hadronization effects (not
included in the CMS results from MCFM) have an impact of the order of
20\%. Also, it is clear that the choice of scheme (either 4F or 5F)
has a considerable impact at the parton-shower NLO QCD level. For example, it
is clear that 5F results compare better with single-$b$-jet
measurements, while 4F results compare best to two-$b$-jet
measurements.

Overall, at the level of total rates the $Z/\gamma^*+b$-jet signatures
have good agreement between theory and experiments, at least when
comparing to the proper ordering of the perturbative expansion (4F vs
5F). New studies with larger experimental data sets will help further
testing this agreement, with particular attention to certain features
that appear over phase space as we will show in the following
subsection.

\begin{figure}[ht]
\begin{center}
\includegraphics[scale=0.3]{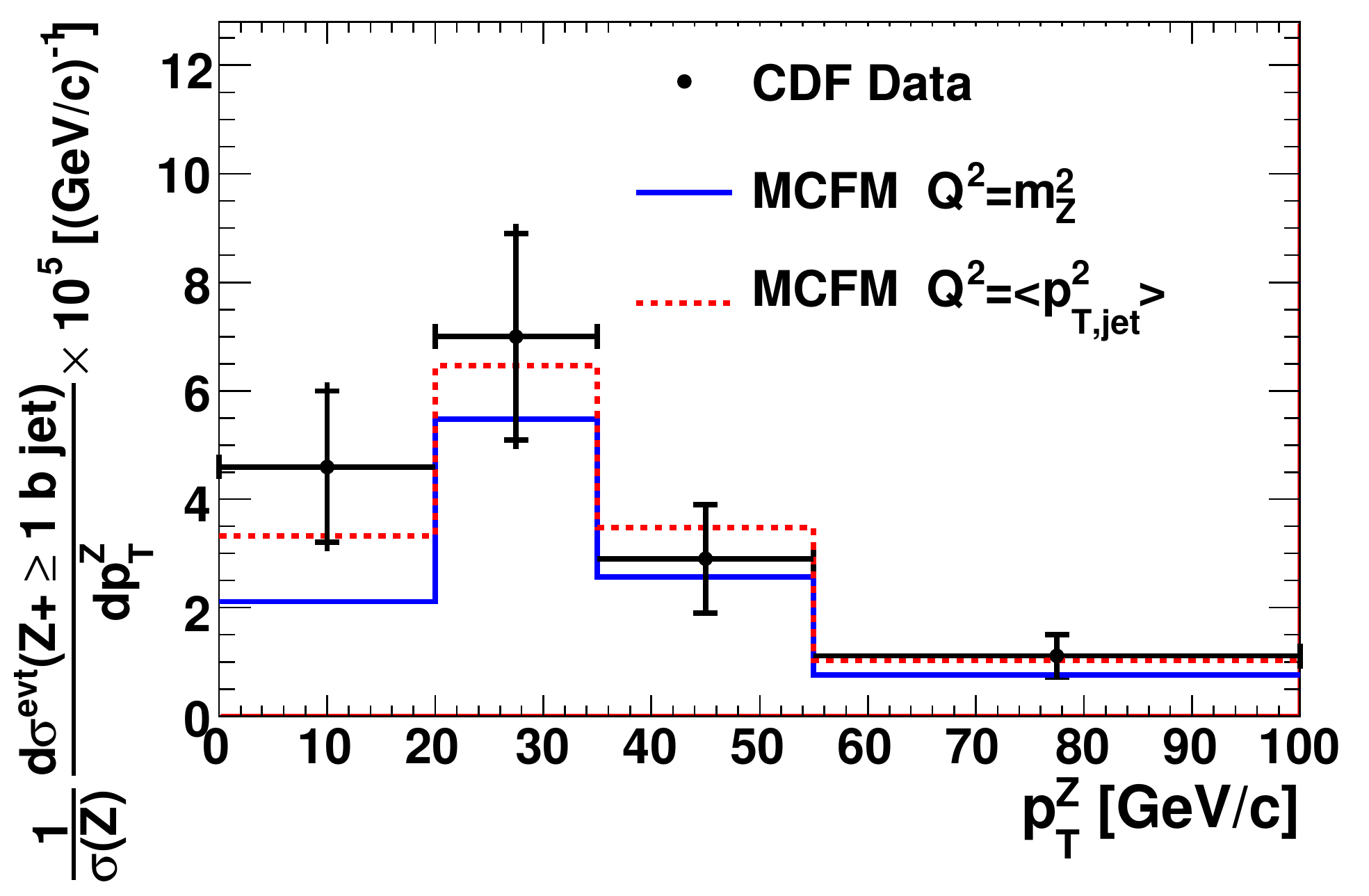}
\hspace{2mm}
\includegraphics[scale=0.31]{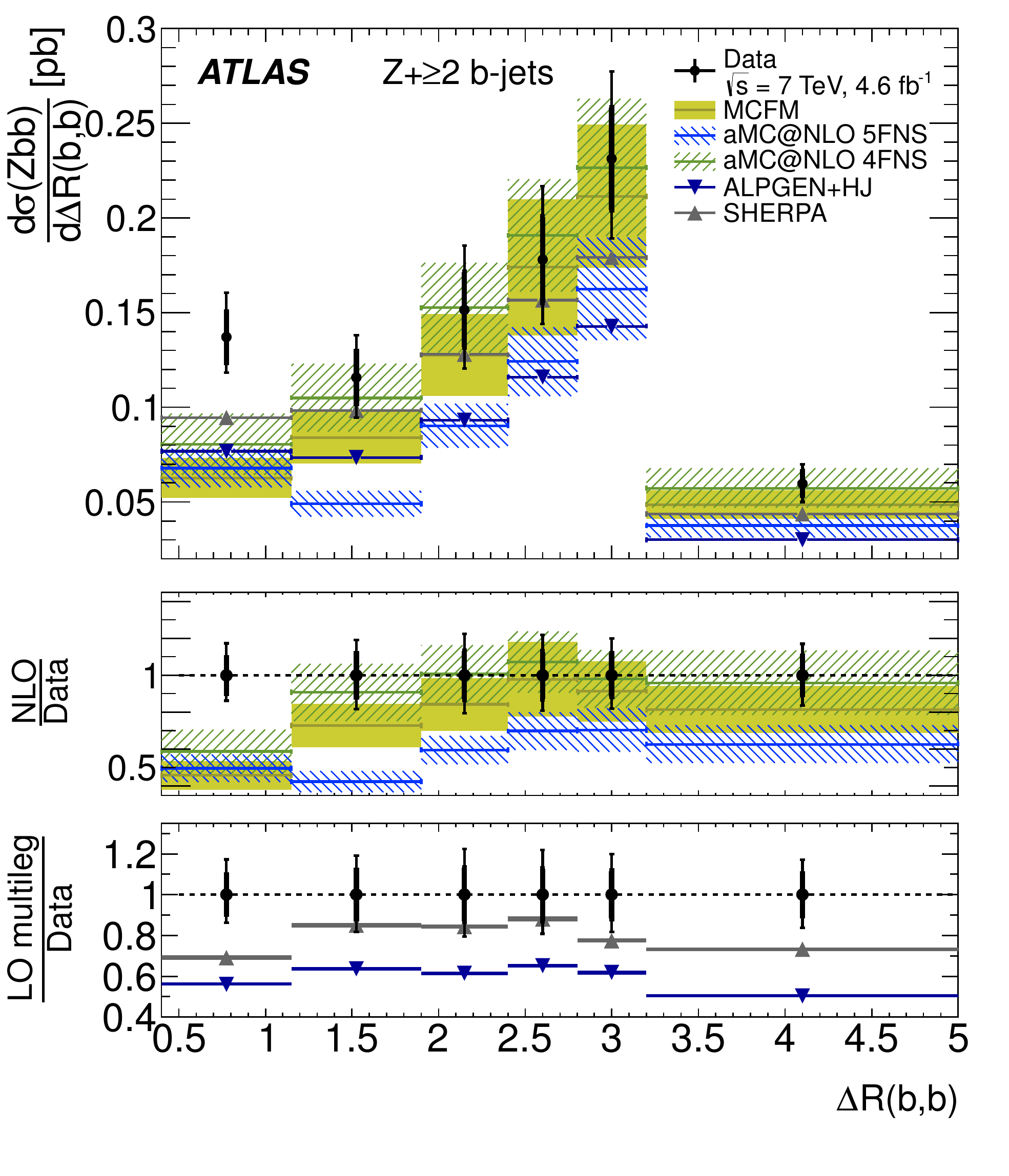}
\includegraphics[scale=0.3]{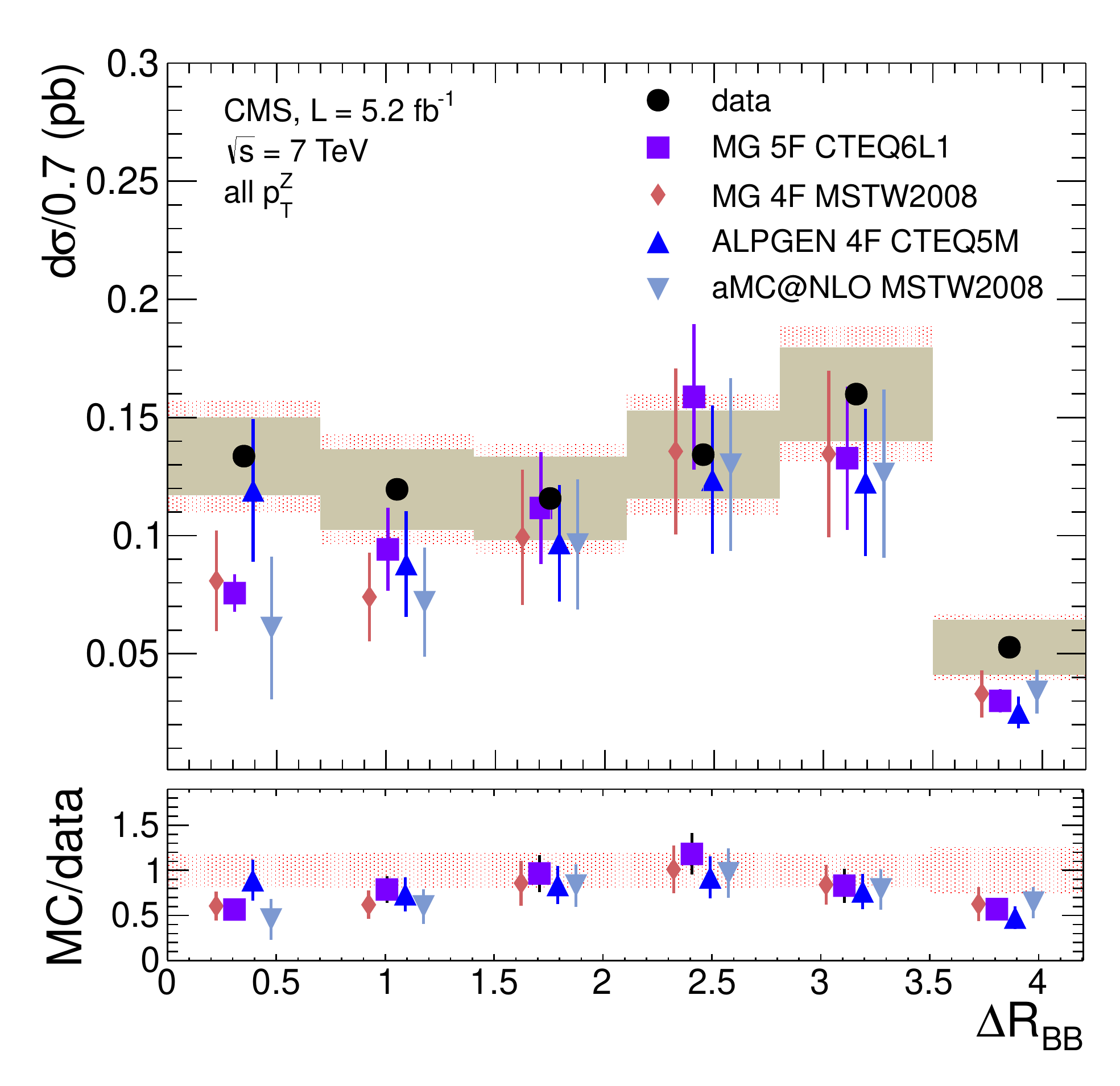}
\hspace{2mm}
\includegraphics[scale=0.3]{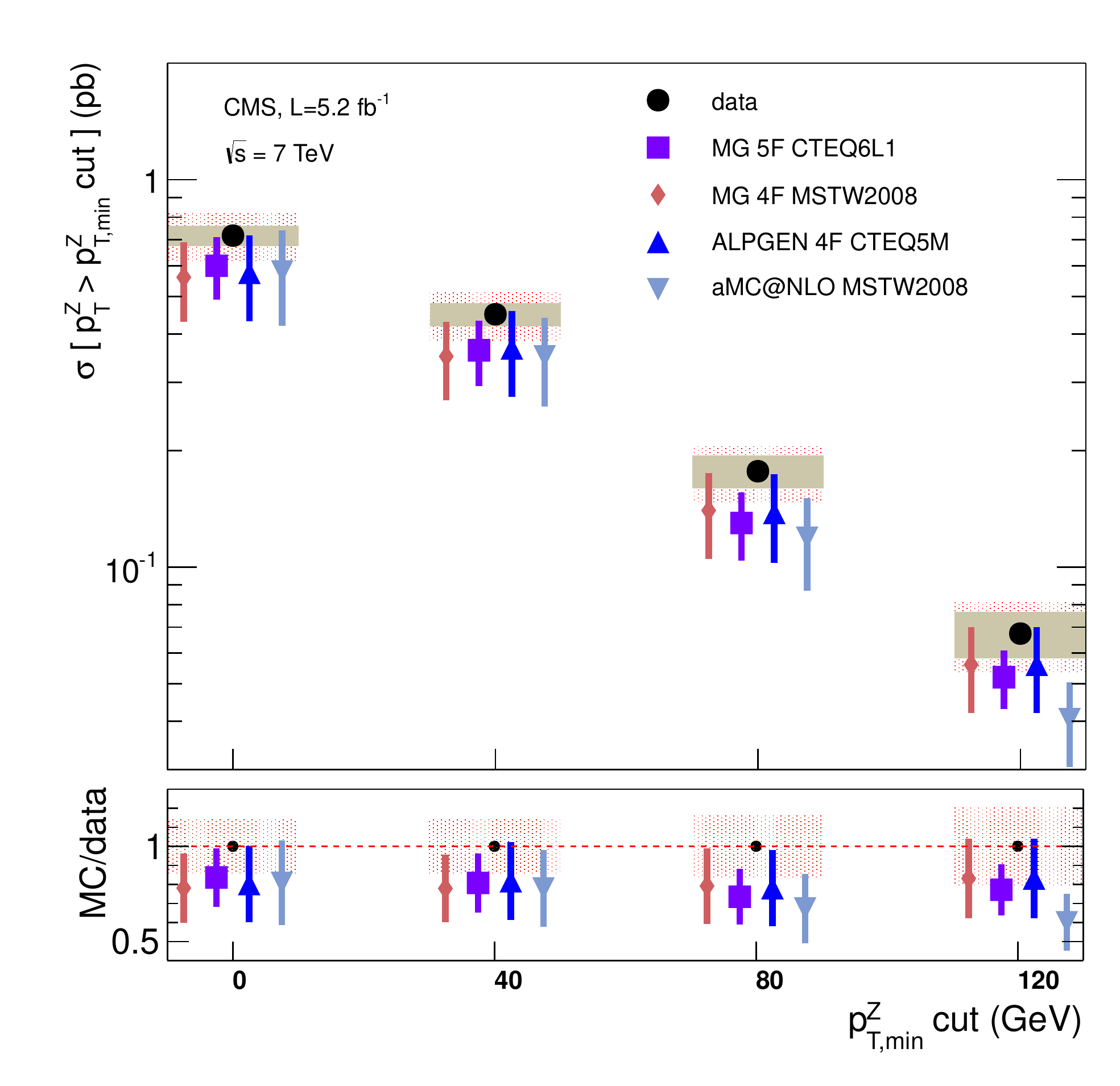}
\end{center}
\caption{Differential cross sections from CDF~\cite{Aaltonen:2008mt} (as a function
of the transverse momentum of the $Z/\gamma^*$ boson),
ATLAS~\cite{Aad:2014dvb} (as a function of the $R$ separation between the two
$b$ jets) and CMS~\cite{Chatrchyan:2013zja} (as a function of the $R$ separation
of the $B$ hadrons and of the transverse momentum of the $Z/\gamma^*$
boson) for various $Z/\gamma^*+b$-jet
signatures compared to theoretical predictions.}
\label{fig:Zb-dXS}
\end{figure}

\subsubsection{Differential cross sections}

In Figure~\ref{fig:Zb-dXS} we have included several plots of
differential cross-section measurements. On the top left, we show a
plot from CDF~\cite{Aaltonen:2008mt} for $Z/\gamma^*+b$-jet$+X$
production as a function of the transverse momentum of the lepton pair
(labeled as $p_{\rm T}^{Z}$). On the top right we show a plot from
ATLAS~\cite{Aad:2014dvb} for $Z/\gamma^*+2\ b$-jet$+X$ production as a
function of the $R$ separation between the $b$ jets. Finally, on the
bottom, we show two plots from CMS as published in
Ref.~\refcite{Chatrchyan:2013zja}. As CMS~\cite{Chatrchyan:2014dha}
did not show differential cross section, we have chosen to show the
ones in CMS~\cite{Chatrchyan:2013zja}, which is a study of the
associated production of a $Z$ boson and two $B$ mesons.  We do not
include plots from D0~\cite{Abazov:2013uza} because they only show
ratios to $Z/\gamma^*+$light jet production.

In general the agreement between theory and experiments on all plots in
Figure~\ref{fig:Zb-dXS} is relatively good. In the ATLAS plot for
$Z/\gamma^*+2b+X$ we notice the poor agreement between the 5F
predictions and data, confirming what previously observed for the
total cross section, i.e. that 5F is not expected to provide reliable
results for processes with two or more $b$ jets (notice that from a
parton-shower point of view, this observable is computed at LO).  Two
of the plots in Figure~\ref{fig:Zb-dXS} show distributions for
$\Delta R$ (for $b$-jets in the case of ATLAS, and $B$ mesons in the
case of CMS) and they both indicate that theoretical descriptions fail
to describe the lower bins.  A failure to describe the small-angle
limit for the production of $b$ jets (hadrons) could be ascribed to
badly described gluon splitting into a slightly collinear $b$-quark
pair. One could then correlate these discrepancies with more
collinear $g\rightarrow b\bar b$ splitting (producing $(bb)$ jets)
which should show up in the tail of distributions in $p_{\rm T}^{Z}$
(and also $p_{\rm T}^{b\rm \ jet}$). Although statistics is poor,
there is no evidence of this in the bottom right plot in
Figure~\ref{fig:Zb-dXS}. Future studies might give more hints of
correlations between discrepancies for low-angle distributions and
large-$p_{\rm T}$ observables. In the end, this might point to a poor
control of gluon splitting to massive quarks and require better
theoretical modeling of $(bb)$ jets.

\subsection{Measurements of $\gamma$ hadronic production
  in association with $b$ jets}
\label{subsec:gammab}

Finally in this subsection we present measurements of the production of a
photon $\gamma$ in association with $b$ jets. Unlike the analyses of the
previous section, where the gauge bosons are tagged through their
decay to leptons, the photon signatures are recorded directly through activity
in electromagnetic calorimeters. Pollution from signatures that mimic photons,
like for example neutral pions decaying into pairs of photons in the
electromagnetic calorimeter, have to be dealt with. We show next the
measurements presented in the literature by CDF~\cite{Aaltonen:2013coa}
and D0~\cite{Abazov:2014hoa}.


\begin{table}[h]
\tbl{$\gamma+b$ jets: signatures, kinematics.}
{\begin{tabular}{@{}c|c|c@{}}\toprule
  \hline
  Experiment & Signatures & Kinematics \\
  \hline
 CDF~\cite{Aaltonen:2013coa}, $1.96$ TeV, $9.1\ {\rm fb}^{-1}$  & 
  $\gamma+b+X$ & 
  \begin{tabular}{l} $E_{\rm T}^\gamma>30$ GeV, $|y^\gamma|<1.04$ \\ $\gamma$
		cone isolation ($R=0.4$):\\
		$E_{\rm T}^{iso}<2.0$ GeV \\ \hline {\it Jets:}\\ JETCLU alg.,
		$R=0.4$\\ 
		$p_{\rm T}^j>20$ GeV, $|y^j|<1.5$\\ $R_{j,\gamma}>0.4$ \end{tabular}  \\
  \hline
  \hline
  D0~\cite{Abazov:2014hoa}, $1.96$ TeV, $8.7\ {\rm fb}^{-1}$  & 
  \begin{tabular}{l} $\gamma+b+X$\\ $\gamma+b+b+X$ \end{tabular} & 
  \begin{tabular}{l} $p_{\rm T}^\gamma>30$ GeV, $|y^\gamma|<1$ \\
		or $1.5<|y^\gamma|<2.5$ \\ 
                $\gamma$ cone isolation ($R=0.4$):\\
		$E_{\rm T}^{iso}<2.5$ GeV \\ \hline {\it Jets:}\\ Midpoint (Run
		II) alg., $R=0.5$\\ 
		$p_{\rm T}^j>15$ GeV, $|y^j|<1.5$ \end{tabular}  \\
  \hline
\hline
\end{tabular}
\label{table:gammab-sign-kin}}
\end{table}

\subsubsection{Experimental setups and differential cross sections}

In Table~\ref{table:gammab-sign-kin} we present the experimental
setups of the $\gamma+b$-jet measurements.  Notice that both CDF and
DO employ a cone isolation criterion to characterize the (prompt)
photon. CDF has considered only the case of central photons
($|y^\gamma|<1.4$), while D0 has presented results including both
central photons ($|y^\gamma|<1$) and forward photons
($1.5<|y^\gamma|<2.5$).

\begin{figure}[ht]
\begin{center}
\includegraphics[scale=0.8]{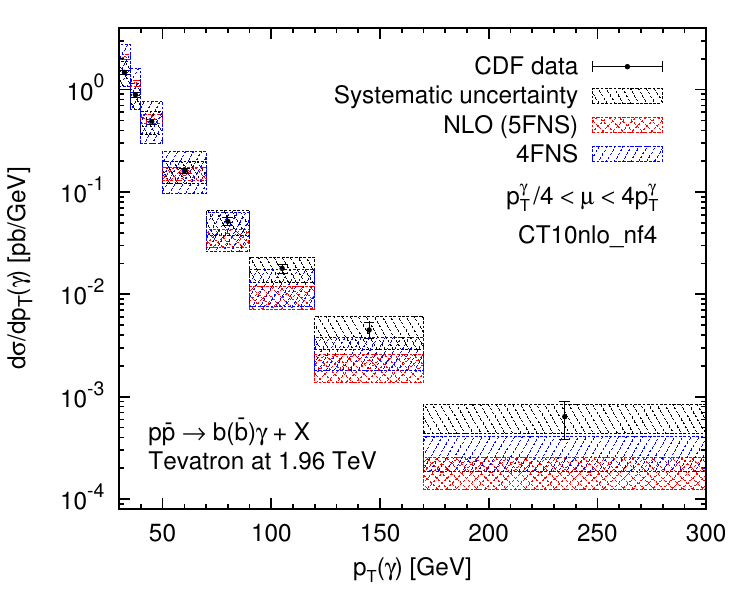}
\end{center}
\caption{Differential cross section as a function of
$p_{\rm T}^\gamma$ measured by CDF~\cite{Aaltonen:2013coa} for
$\gamma+b$-jet production  compared to theoretical
predictions~\cite{Hartanto:2013aha}.}
\label{fig:gammab-dXSCDF}
\end{figure}

\begin{figure}[ht]
\begin{center}
\includegraphics[scale=0.8]{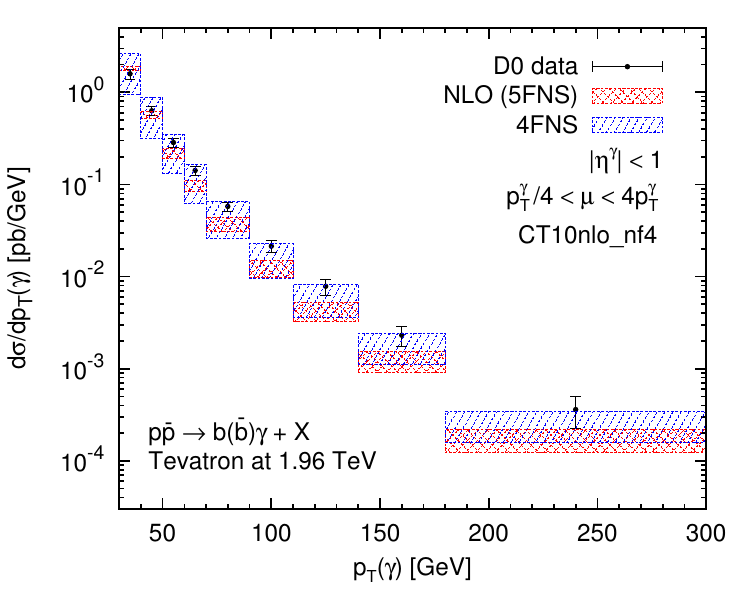}
\includegraphics[scale=0.8]{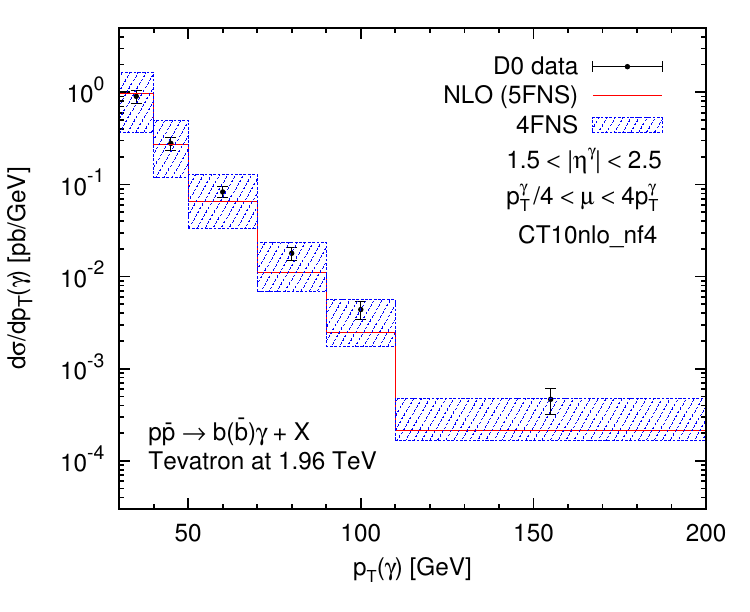}
\end{center}
\caption{Differential cross sections as a function of
$p_{\rm T}^\gamma$ measured by D0~\cite{Abazov:2012ea} for
$\gamma+b$-jet production compared to theoretical
predictions~\cite{Hartanto:2013aha}.}
\label{fig:gammab-dXSD0}
\end{figure}

\begin{figure}[ht]
\begin{center}
\includegraphics[scale=0.3]{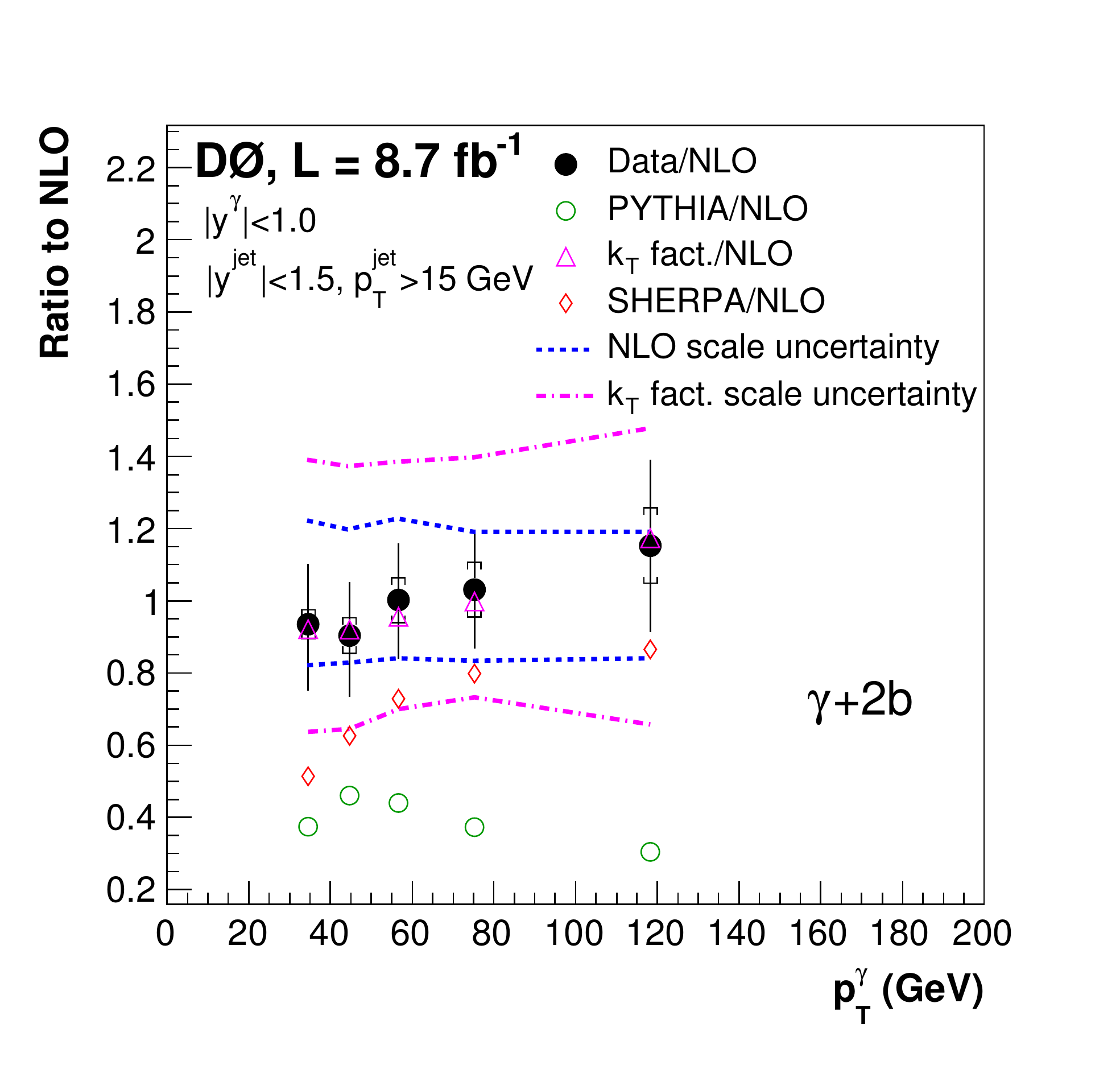}
\includegraphics[scale=0.3]{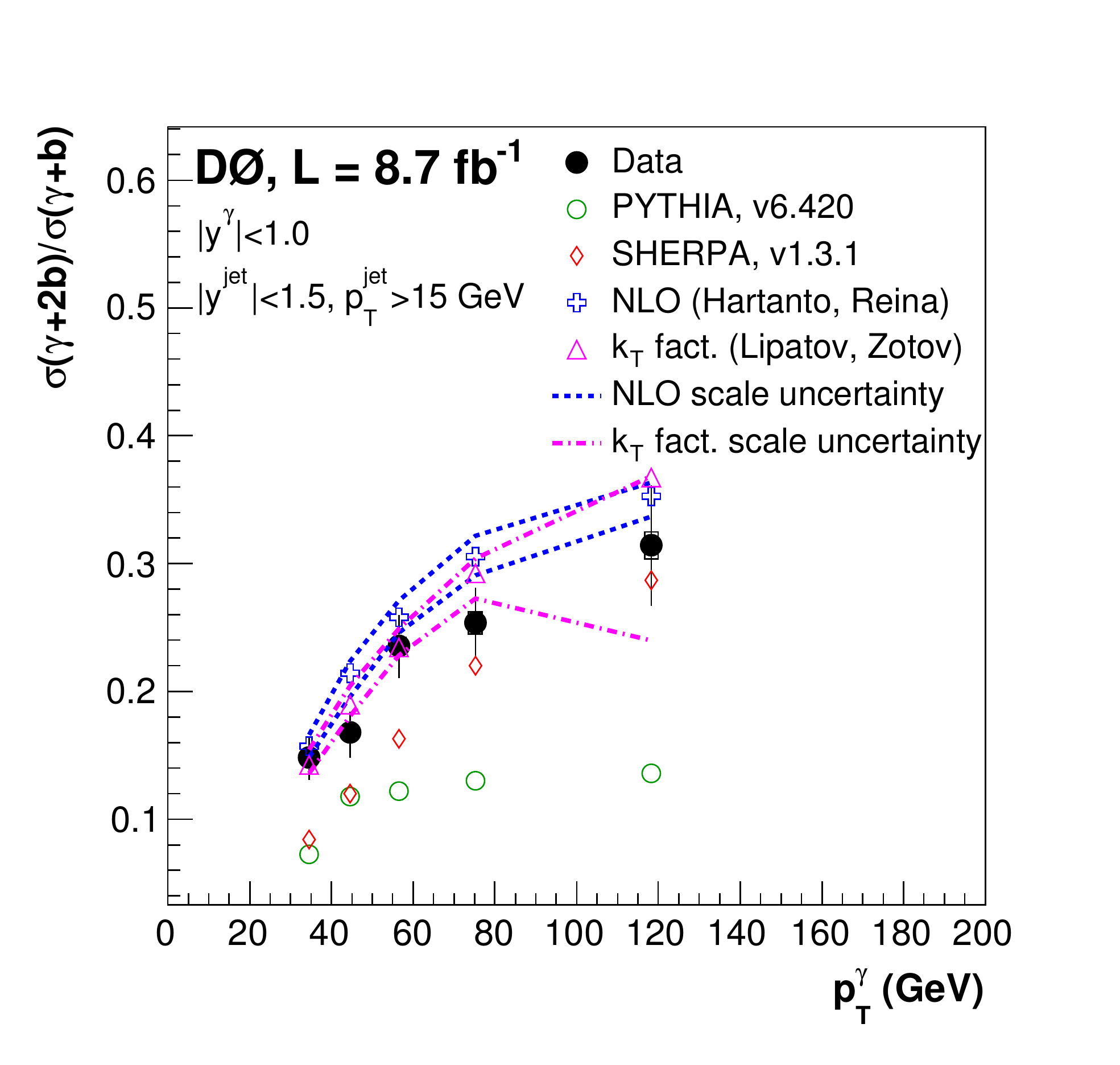}
\end{center}
\caption{Ratio of D0~\cite{Abazov:2014hoa} data (and other theory
  predictions) to NLO QCD prediction (left) for $\gamma+b+b+X$
  production, and ratio $(\gamma+2b+X)/(\gamma+b+X)$ compared to
  theoretical predictions (right), as a function of $p_{\rm
    T}^\gamma$}.
\label{fig:gammabb-dXSD0}
\end{figure}

Both CDF and D0 have shown their results differentially as a
distribution in the transverse momentum of the photon ($p_{\rm
  T}^\gamma$).  We cannot therefore give a table of results, but will
illustrate the comparison between experimental measurements and theory
through the $p_{\rm T}^\gamma$ distributions themselves. The
comparison with the $\gamma+b+X$ measurement of
CDF~\cite{Aaltonen:2013coa} is illustrated in
Figure~\ref{fig:gammab-dXSCDF} while the comparison with both the
$\gamma+b+X$ and the $\gamma+b+b+X$ measurements of
D0~\cite{Abazov:2012ea,Abazov:2014hoa} is illustrated in
Figures~\ref{fig:gammab-dXSD0} and \ref{fig:gammabb-dXSD0}.  For the
comparison with the $\gamma+b+X$ measurements of both
CDF~\cite{Aaltonen:2013coa} and D0~\cite{Abazov:2012ea} we show plots
from Ref. ~\refcite{Hartanto:2013aha} since they compare to the most
up to date theoretical calculations, while for the comparison with the
$\gamma+b+b+X$ measurements of D0~\cite{Abazov:2014hoa} we present
plots taken from the experimental paper itself where results from
Ref. ~\refcite{Hartanto:2013aha} have been used. In
Figure~\ref{fig:gammabb-dXSD0} data are also compared to results from
a LO calculation with $p_T$-dependent
PDF~\cite{Lipatov:2005wk,Lipatov:2012rg}, where however the
uncertainty of the prediction is difficult to estimate and to compare
with a full NLO calculation using integrated PDF.


It can be appreciated that theory prediction in general describe the
data. This is a case even for $\gamma+1b$ jet, where the 4F prediction
is in better agreement with data than the 5F prediction since,
as explained in Section~\ref{sec:theory}, the dominant production mode
at the Tevatron is via the $q\bar{q}$ channel, which is not influenced
by large initial-state logarithms.  Still, theory distributions seem
slightly softer and fall a little below data at high $p_{\rm
  T}^\gamma$. Measurements at the LHC will help pushing distributions
to much higher $p_{\rm T}^\gamma$ and confirm or not these trends.  In
general there is considerable overlap between 4F and 5F predictions
(denoted in the plots by 4FNS and 5FNS) for 1 $b$-jet observables. As
one might expect, from the missing extra $b$ quark, 5F predictions
tend to drop faster as a function of the photon transverse momentum.

\section{Conclusions}
\label{sec:conclusions}

In this review we have presented the current status of electroweak
vector-boson associated hadronic production with
$b$ jets. Measurements and theoretical predictions for $W+b$-jet,
$Z/\gamma^*+b$-jet, and $\gamma+b$-jet signatures were thoroughly
reviewed. Results from the major high-energy-physics experimental
collaborations, CDF and D0 at the Tevatron, and ATLAS and CMS at the
LHC, were summarized.

By now we have reached a mature level in the study of these processes.
This is very important given the impact that related studies can
have in constraining couplings of the Higgs boson to $b$ quarks,
direct studies of heavy-quark parton distribution functions, and
searches for new physics.

Although in general we find good agreement between theoretical
predictions and experimental measurements, some tensions are still
present and important challenges are left for the near future.  With
the larger data sets collected at $\sqrt{s}=8$ TeV at the LHC, and
with the even larger ones expected at and above $\sqrt{s}=13$ TeV, we
expect improved statistical errors as well as reduced systematic
errors (thanks to a better understanding of the detectors). Matching
these reduced errors will be challenging for theory predictions. Full
parton-showered results, including NLO-QCD corrections and merging
prescriptions for different jet multiplicities, as well as
systematic studies of hadronization and underlying-event models should
become the standard approach. The need for NNLO QCD corrections
as well as the inclusion of EW corrections and resummation of
large logarithmic corrections could play a role in selected
observables and will have to be investigated. More exclusive
studies that include larger amount of heavy jets are now also
theoretically and experimentally approachable and could lead to better
control of the $V+b$ jet processes at hadron colliders.

\section*{Acknowledgments}
The Authors would like to thank most of all J.~ M.~Campbell,
H.~B.~Hartanto, and D. Wackeroth, with whom many of the studies
reviewed in this paper have been completed.  The Authors would
  also like to thank many colleagues on CDF, D0, ATLAS, and CMS, in
  particular Tobias Golling, Chiara Mariotti, Christopher Neu,
  Alexander Nikitenko, and Evelyn Thomson, with whom they have enjoyed
  several fruitful collaborations, and from whom they have learned so
  many important aspects of the processes presented in this review.
The Authors are very grateful to the Aspen Center for Physics for its
kind hospitality and support during part of the work that led to this
review.  F.F.C. has been in part supported by an APS Travel Grant and
by the Simons Foundation.  The work of F.F.C. is supported by the
Alexander von Humboldt Foundation, in the framework of the Sofja
Kovalevskaja Award 2014, endowed by the German Federal Ministry of
Education and Research.  The work of L.R. is partially supported by
the U.S. Department of Energy under Grant DE-FG02-13ER41942.

\bibliographystyle{ws-mpla}

\end{document}